\documentclass[twocolumn, trackchanges]{aastex631}
\usepackage{gensymb}
\usepackage{amsmath}
\usepackage{enumerate}
\usepackage{CJK}
\usepackage{multirow}
\usepackage{epsfig}
\usepackage[para]{threeparttable}

\newcommand{\lya}{\ensuremath{\rm Ly\alpha}}
\newcommand{\Ha}{\ensuremath{\rm H\alpha}}
\newcommand{\Hbeta}{\ensuremath{\rm H\beta}}
\newcommand{\HI}{\ion{H}{1}}
\newcommand{\HII}{\ion{H}{2}}
\newcommand{\NII}{[\ion{N}{2}]}
\newcommand{\CIV}{\ion{C}{4}}
\newcommand{\SiIIfs}{\ion{Si}{2}*}
\newcommand{\SiII}{\ion{Si}{2}}
\newcommand{\OI}{\ion{O}{1}}
\newcommand{\OII}{[\ion{O}{2}]}
\newcommand{\OIII}{[\ion{O}{3}]}
\newcommand{\CII}{\ion{C}{2}}
\newcommand{\SiIV}{\ion{Si}{4}}
\newcommand{\FeII}{\ion{Fe}{2}}
\newcommand{\AlII}{\ion{Al}{2}}

\shorttitle{Gas Accretion in Protocluster Galaxies at $z = 2.3$}
\shortauthors{Bolda et al.}
\graphicspath{{./}{figures/}}
\begin{document}
\begin{CJK*}{UTF8}{gbsn}

\title{Gas Accretion Traced by Blue-Dominated \lya\ Emission and Redshifted UV Absorption Lines in Protocluster Galaxies at $z = 2.3$ from the KBSS-KCWI Survey \footnote{This paper presents findings from observations conducted at the W.M. Keck Observatory, which is operated as a scientific partnership among the California Institute of Technology, the University of California, and the National Aeronautics and Space Administration. The Observatory was made possible by the generous financial support of the W. M. Keck Foundation.} \footnote{The HST/ACS observations presented in this paper are available on the Space Telescope Science Institute's Mikulski Archive for Space Telescopes (MAST) at \dataset[doi: 10.17909/j2mb-9350]{https://doi.org/10.17909/j2mb-9350}.}}

\author[0000-0002-6913-8580]{Claire Bolda}
\affiliation{The Leonard E. Parker Center for Gravitation, Cosmology and Astrophysics, Department of Physics, University of Wisconsin-Milwaukee, 3135 N. Maryland Avenue, Milwaukee, WI 53211, USA}
\affiliation{Department of Physics and Astronomy, Texas A\&M University, College Station, TX, 77843-4242 USA}
\affiliation{George P. and Cynthia Woods Mitchell Institute for Fundamental Physics and Astronomy, Texas A\&M University, College Station, TX, 77843-4242 USA}

\author[0000-0001-5113-7558]{Zhihui Li (李智慧)}
\affiliation{Cahill Center for Astrophysics, California Institute of Technology, MC 249-17, 1200 East California Boulevard, Pasadena, CA 91125, USA}
\affiliation{Center for Astrophysical Sciences, Department of Physics \& Astronomy, Johns Hopkins University, Baltimore, MD 21218, USA}

\author[0000-0001-9714-2758]{Dawn K. Erb}
\affiliation{The Leonard E. Parker Center for Gravitation, Cosmology and Astrophysics, Department of Physics, University of Wisconsin-Milwaukee, 3135 N. Maryland Avenue, Milwaukee, WI 53211, USA}

\author[0000-0002-4834-7260]{Charles C. Steidel}
\affiliation{Cahill Center for Astrophysics, California Institute of Technology, MC 249-17, 1200 East California Boulevard, Pasadena, CA 91125, USA}

\author[0000-0003-4520-5395]{Yuguang Chen (陈昱光)}
\affiliation{Department of Physics \& Astronomy, University of California, Davis, 1 Shields Ave., Davis, CA 95616, USA}

\correspondingauthor{Claire Bolda}
\email{boldaclaire@tamu.edu}

\begin{abstract}

\lya\ emission with a dominant blueshifted peak can probe gas flowing through the circumgalactic medium as it accretes onto galaxies and fuels new star formation, although it has seldom actually been observed. Here we present new Keck Cosmic Web Imager observations of the extended \lya\ halos surrounding Q1700-BX710 and Q1700-BX711, a pair of UV continuum-selected Keck Baryonic Structure Survey (KBSS) galaxies at $z = 2.3$ in the HS1700+643 protocluster. We find that BX710 and BX711's \lya\ halos are aligned with a large-scale galaxy filament consisting of thirteen spectroscopically identified protocluster galaxies. By measuring the peak separation and blue-to-red peak flux ratio of the \lya\ emission profiles throughout these galaxies' \lya\ halos, we have obtained measurements of their spatially varying velocity structure. The prevalence of blue-dominated \lya\ emission profiles throughout BX711's \lya\ halo suggests actively accreting gas. We fit a clumpy, multiphase Monte Carlo Radiative Transfer model which assumes a radially varying clump velocity to the spatially resolved \lya\ emission throughout BX710 and BX711's \lya\ halos and simultaneously fit these galaxies' average down-the-barrel UV absorption profile with a radially varying velocity model. The results of these models are consistent with a combination of \HI\ and higher-metallicity gas accretion for both galaxies, especially BX711, which exhibits inflow-driven kinematics throughout most of its \lya\ halo. We consider various accretion scenarios to explain these findings, including accretion of metal-enriched gas from the cosmic web, galaxy interactions, and recycled gas from the circumgalactic medium, all of which are compatible with our current observations.

\end{abstract}


\section{Introduction} \label{sec:intro}

    According to cosmological simulations, galaxies across cosmic time acquire fuel to form new stars by accreting gas from various environmental sources, often simultaneously \citep{2005MNRAS.363....2K, L'Huillier_2012, Hafen_2019, Mitchell_2021}. Some of these environmental sources include the cosmic web \citep{2005MNRAS.363....2K, Dekel_2006}, the circumgalactic medium (CGM)---a gaseous expanse surrounding each galaxy that acts as a gateway through which gas enters and exits \citep{Oppenheimer_Dave_2008, Tumlinson_2017, Faucher-Giguere_2023}---satellite galaxies \citep{Hafen_2019, Mitchell_2021}, and galaxy mergers \citep{Silk_2013}.

    Flowing along the cosmic web are cool filaments of \HI\ gas which are pulled through the CGM due to the influence of gravity and subsequently accrete onto galaxies at cosmic noon ($z \sim 2-3$) and earlier times \citep{Birnboim_Dekel_2003, 2005MNRAS.363....2K, Dekel_2009}. Low metallicity gas from the cosmic web is predicted to be the primary contributor of fresh, new fuel for star formation, and consequently of growth, in galaxies \citep{2005MNRAS.363....2K, Dekel_2009, Hafen_2019}.

    In the case of the CGM, stellar feedback propels metal-enriched gas out of a galaxy's interstellar medium (ISM) and into this vast gas interchange \citep{Veilleux_2005, Erb_2015, Fraternali_2017, Tumlinson_2017}. If the ejected gas does not have a high enough velocity to escape the galaxy, the CGM will store it indefinitely unless the influence of gravity compels it to fall in towards the galaxy, where it may subsequently be recycled to form new stars \citep{Oppenheimer_2010, Angles_2017}. While this higher metallicity gas does provide galaxies with fuel for star formation, it does not increase their total mass. 
    
    Metal enriched gas may be driven out of satellite galaxies and then be accreted onto the host galaxies they orbit as a result of stellar feedback or stripping \citep{Angles_2017, Hafen_2019}. Although this gas likewise has a higher metallicity than \HI\ gas from the cosmic web, in this case it helps elevate both the stellar mass and total mass of the host galaxies \citep{Hafen_2019}.  

    Interactions between galaxies and their environments in the form of both inflowing and outflowing gas can be detected via the $\lambda 1215.67$ \AA \ \lya\ emission line, which reveals extended halos of \HI\ gas (\lya\ halos) within the CGM of many galaxies at both low and high redshifts \citep[e.g.,][]{Steidel_2011, Matsuda_2012, Ostlin_2014, Hayes_2015, Stark_2016, Wistotzki_2016, Leclercq_2017, Kusakabe_2022}. Additionally, various other absorption, emission, and P-Cygni features in galaxies' rest frame UV spectra act as indicators of metal-enriched gas exchanges among galaxies, the CGM, and the IGM (e.g.,\ \citealt{1993ApJS...86....5K, 2003ApJ...588...65S, Jones_2012, Marques-Chavez_2020}).

    Observational evidence in support of galaxy growth via gas accretion at cosmic noon ($z\sim2-3$) and higher redshifts remains limited due to challenges such as low covering fractions of accreting \HI\ gas filaments. Currently, detections of redshifted metal absorption lines in galaxy spectra, which can indicate metal-enriched gas accreting onto a galaxy, comprise the majority of this evidence \citep[e.g.,][]{Giavalisco_2011, 2012ApJ...760..127M, Rubin_2012, Calabro_2022, Weldon_2023}. Despite their relative abundance, it can be difficult to disentangle whether redshifted absorption lines result from recycled accretion, metal-enriched gas from the IGM, or even a combination of cool gas accretion from the cosmic web and recycled accretion, and consequently to determine which environmental sources may be responsible for the accreting gas they trace \citep[e.g.][]{Giavalisco_2011, 2012ApJ...760..127M, Rubin_2012, Calabro_2022, Weldon_2023}. Indications of accreting gas from \lya\ emission, although less prevalent, can be advantageous to search for in light of this since \lya\ emission is a much stronger feature \citep{Hayes_2015}, is less dependent on metallicity, and can trace accreting \HI\ gas from the cosmic web more directly than redshifted absorption lines. 
     
    Proposed origins of the \lya\ photons which compose \lya\ halos surrounding star-forming galaxies include central \HII\ regions in galaxies and cool, inflowing \HI\ gas streams in the CGM. In the first scenario, \lya\ photons are produced by the photoionization and Case B recombination of hydrogen atoms in a galaxy's central \HII\ regions and then scatter outwards into the CGM \citep[e.g.][]{Partridge_Peebles_1967, Charlot_1991, Gould_Weinberg_1996}. In the latter scenario, collisional excitation in inflowing streams of \HI\ gas leads to cooling radiation in the form of \lya\ photons that travel throughout the CGM without a preferential direction \citep{Haiman_2000, Dijkstra_2006a, Dijkstra_2006b, Faucher_2011, Cantalupo_2017}. 
    
    \lya\ photons may undergo resonant scattering by \HI\ gas in the CGM, which can be static or have a bulk radial inflow or outflow velocity. Each time a \lya\ photon scatters resonantly off of a \HI\ atom, the thermal velocity of this atom, in combination with the bulk velocity of the CGM gas, shifts the frequency of the \lya\ photon slightly blueward or redward of its rest wavelength \citep{Dijkstra_2006a, Verhamme_2006}. Depending on the CGM's \HI\ column density, multiple scattering events may occur before \lya\ photons receive an adequately large velocity shift to escape the CGM without further interactions \citep[e.g.,][]{Dijkstra_2006a, Steidel_2010}. The longer a \lya\ travels through a galaxy, the more likely it is to be extinguished by dust \citep[e.g.,][]{Charlot_1991}.
    
    In an outflowing CGM, \lya\ photons typically escape after scattering off of \HI\ gas on the opposite side of the galaxy from the observer, during which process they receive a substantial frequency redshift \citep{Zheng_2002, Adelberger_2003, Verhamme_2006, Steidel_2010, Erb_2015}.
    
    Alternately, as \lya\ photons scatter off of \HI\ gas in an inflowing CGM, they receive a sizeable frequency blueshift as well as an additional frequency shift from its thermal and bulk velocities \citep{Dijkstra_2006a, Zheng_2002}. As a result, \lya\ photons in an inflowing CGM usually escape after scattering off of \HI\ gas on the closest side of the galaxy to the observer \citep{Dijkstra_2006a, Zheng_2002}.
    
    Since resonant scattering of \lya\ photons off of inflowing gas can produce dominant blueshifted \lya\ emission peaks, the presence of strong blueshifted peaks in a galaxy's \lya\ profile is often expected to indicate that it is accreting gas \citep{Dijkstra_2006a, Dijkstra_2006b, Verhamme_2006}. Dominant blueshifted \lya\ peaks produced by other \lya\ emission mechanisms such as fluorescence from AGN and starbursts may illuminate accreting gas as well \citep{Vanzella_2017}. Across cosmic time, nearly all Lyman-$\alpha$ Emitters (LAEs), Lyman-break galaxies (LBGs), and UV color-selected galaxies which exhibit \lya\ emission profiles feature dominant redshifted peaks and far less prominent or undetectable blueshifted peaks in their spectra, suggesting that gas is flowing outward from these galaxies into the CGM \citep[e.g.,][]{Pettini_2001, 2003ApJ...588...65S, Verhamme_2008, Hayes_2015, Martin_2015, Leclercq_2017, Erb_2023}. Blueshifted metal absorption lines are often observed in tandem with these red-dominated \lya\ emission profiles \citep[e.g.,][]{2003ApJ...588...65S, Erb_2023}.
    
    However, a couple of exceptions have been reported by \citet{Kulas_2012}, who found \lya\ profiles with dominant blueshifted peaks and significantly less prominent redshifted peaks in the spectra of two out of 18 UV continuum-selected galaxies, Q2346-BX181 at $z = 2.5$ and Q1549-C20 at $z=3.1$. Likewise, \citet{Trainor_2015}, \citet{Berg_2018}, and \citet{Marques_Chaves_2022} discovered \lya\ profiles with stronger blueshifted peaks in the spectra of 33 out of 129 total LAEs with more than one \lya\ peak at $z \gtrsim 2.5-3$; a lensed LAE at $z = 1.8$; and an LAE at $z = 3.6$, respectively. In the Keck Baryonic Structure Survey \citep[KBSS;][]{Rudie_2012, Steidel_2014, Strom_2017, Theios_2019, Chen_2021, Erb_2023}, of which this paper is part, only 4.3\% of sample galaxies at $z > 2$ with confirmed redshifts from both nebular emission lines and \lya\ emission, or 21 total galaxies, satisfy the criterion $z_{\lya} < z_{neb}$, and thereby have blue-dominated \lya\ profiles. Intriguingly, stronger blueshifted \lya\ peaks have not yet been reported among potentially interacting galaxy pairs with spectra that feature \lya\ emission \citep[e.g.,][]{Cooke_2009}.
    
    More recently, integral field spectroscopy has enabled closer examination of the \lya\ profiles throughout \lya\ halos, as well as \lya\ blobs, which are regions of extremely bright \lya\ emission spanning hundreds of kiloparsecs. Integral field observations have  revealed noticeable spatial variations in the spectral parameters of \lya\ profiles, including their asymmetry, intensity, blue-to-red peak flux ratio, and peak separation \citep{Erb_2018, Claeyssens_2019, Leclercq_2020, Erb_2023, Furtak_2022, Guo_2023, Mukherjee_2023}. The blue-to-red peak flux ratio can especially help determine whether and to what extent inflowing \HI\ gas is present in the CGM \citep{Dijkstra_2006b}. 
    
    With the aid of integral field observations, several reports have disclosed the presence of blue-dominated \lya\ emission throughout some \lya\ blobs \citep{Ao_2020, Daddi_2021, Li_2022}. In all cases, these reports have inferred that the most probable source of this highly unusual stronger blueshifted \lya\ peak is gas flowing into these galaxies from environmental sources. 
    
    Likewise, integral field observations of \lya\ halos associated with individual galaxies have demonstrated that blue-dominated \lya\ profiles can exist in \lya\ halos in addition to the more commonly observed, red-dominated \lya\ profiles. Notably, \citet{Furtak_2022} presented integral field observations of a gravitationally lensed LAE at $z = 3.2$, RXC0018-LAE1.2, which revealed slightly blue-dominated, double-peaked \lya\ emission in the central regions of this LAE's \lya\ halo and strongly blue-dominated emission throughout most of its outer regions. \citet{Furtak_2022} concluded that inflowing gas from an undetermined source is likely responsible for the observed blue-dominated \lya\ emission in this LAE's \lya\ halo.
    
    In addition, \citet{Mukherjee_2023} reported integral field observations of blue-dominated \lya\ emission in \lya\ halos associated with three LAEs at z$\sim$3-5. These halos primarily feature strongly blue-dominated \lya\ emission in their inner regions and red-dominated \lya\ emission at their outer edges, although one of the LAEs, which is located at $z = 3.6$, features two spatially distinct patches of blue-dominated \lya\ emission separated by red-dominated \lya\ emission \citep{Mukherjee_2023}. Given the prevalence of blue-dominated \lya\ profiles in these halos, \citet{Mukherjee_2023} inferred that the LAEs associated with these halos may be accreting gas from an unknown source. 
    
    Because accreting gas can be very difficult to observe, more integral field observations of \lya\ halos surrounding galaxies with blue-dominated \lya\ profiles may, in combination with \lya\ radiative transfer modeling, facilitate the discovery of individual galaxies which are actively accreting gas and potentially even the environmental sources of this gas. In an attempt to help fill this gap, we present spatially resolved integral field observations of the \lya\ halos encompassing a pair of UV continuum-selected galaxies, Q1700-BX710 and Q1700-BX711, in the HS1700+643 protocluster at $z = 2.300 \pm 0.015$ \citep{Steidel_2005} using the Keck Cosmic Web Imager (KCWI) as part of the KBSS-KCWI survey \citep{Chen_2021}. BX711 features blue-dominated \lya\ emission in both its star-forming regions and its extended \lya\ halo, which indicates that it may be accreting gas \citep{Dijkstra_2006a}. Here we investigate the extent to which gas may be falling into either galaxy, and the environmental sources from which it may originate, by examining the spatially varying \lya\ profiles in these galaxies' \lya\ halos and their rest-frame UV metal absorption features. Due to the rarity of observed \lya\ halos with blue-dominated \lya\ emission and a lack of environmental information in most cases, this galaxy pair presents an excellent opportunity to study the impact of environment on galaxy growth.

    In this paper, we describe our target selection, observations, and data reduction in Section \ref{sec:obs}. We detail the spatial measurements we performed on BX710 and BX711's \lya\ halos and  galaxy filaments in the HS1700+64 protocluster in Section \ref{sec:sm}. Then we recount our analysis of the \lya\ profiles throughout BX710 and BX711's \lya\ halos, as well as their 1D rest-frame UV spectra, and present our findings in Section \ref{sec:spectra}. In Section \ref{sec:modeling}, we describe the application of radiative transfer models to the \lya\ profiles and models to select metal UV absorption features. Finally, we summarize our results and discuss potential inflow scenarios in Section \ref{sec:dis}. Here we assume a $\Lambda$CDM cosmology with $\Omega_m = 0.31$, $\Omega_{\Lambda} = 0.69$, and $H_0 = 67.7$ kms$^{-1}$Mpc$^{-1}$ \citep{Planck_2018}. At the mean systemic redshift of BX710 and BX711, an angular separation of $1.0\arcsec$ \ corresponds to a distance of $8.4$ proper kpc.

\section{Targets, Observations, and Data Reduction} \label{sec:obs}

\subsection{Targets and Additional Objects in the Field}
\label{subsec:targetsandfriends}

Our targets consist of the UV continuum-selected galaxy pair Q1700-BX710 and Q1700-BX711, which are included in the Keck Baryonic Structure Survey \citep[KBSS;][]{Rudie_2012, Steidel_2014, Strom_2017, Theios_2019, Chen_2021, Erb_2023}. BX710 is an $L_{*}$ star-forming galaxy with a stellar mass of $M_{*} = 2.2 \times 10^{10} \ M_{\odot}$ at $z_{\rm sys} = 2.2946$, while BX711 is a comparatively young, low-mass ($M_{*} = 6.6 \times 10^8 \ M_{\odot}$) star-forming galaxy at $z_{\rm sys} = 2.2947$ \citep{Reddy_2009, Steidel_2014}. These galaxies are separated by only $6\arcsec$ on the sky, or $48$ projected kpc, and their systemic redshifts determined from nebular emission lines are nearly identical. 

BX711 was initially targeted for KCWI observations as part of the extreme nebular emission line sample in \citet{Erb_2023} since it satisfies the extreme emission line ratio criteria defined by \citet{Erb_2016} as $\log($\NII$/$\Ha$) \leq -1.1$ and $\log($\OIII$/$\Hbeta$) \geq 0.75$. However, because of the blue-dominated \lya\ profile in BX711's rest-frame UV spectrum and its proximity to BX710, which complicates the interpretation of its \lya\ measurements, this galaxy was excluded from \citet{Erb_2023}'s extreme emission line galaxy sample in order to comprehensively study this enigmatic galaxy and its companion separately in this paper.  

Physical and spectral properties of BX710 and BX711 (star formation rate, nebular emission line ratios, etc.) measured by the Keck Baryonic Structure Survey are listed in Table \ref{tab:properties}. The nebular emission line measurements in this table were observed by KBSS using Keck-MOSFIRE \citep[KBSS-MOSFIRE; e.g.,][]{Steidel_2014}. Dust reddening, $E(B-V)$, was determined from the measured \Ha/\Hbeta\ ratio assuming the SMC extinction law. As in \citet{Erb_2023}, this galaxy pair's star formation rates were estimated from their \Ha\ luminosities and the star formation rate calibration from \citet{Theios_2019}, which is based on the BPASSv2.2 stellar population model \citep{Stanway_Eldridge_2018}, after correcting for dust extinction.
\begin{deluxetable}{c | c c}[t]
    \renewcommand{\arraystretch}{1.25}
    \tablecaption{Properties of Q1700-BX710 and Q1700-BX711}
    \label{tab:properties}
    \tablehead{\multicolumn{1}{c|}{Galaxy}&\multicolumn{1}{c}{Q1700-BX710}&\multicolumn{1}{c}{Q1700-BX711}}
    \startdata
    z$_{\rm neb}$ (1) & 2.2946 & 2.2947 \\
    M$_{*}$ (M$_{\odot}$) (2) & 2.2 $\times$ 10$^{10}$ & 6.6 $\times$ 10$^{8}$ \\
    $F_{\rm H\alpha}$ (10$^{-17}$ erg s$^{-1}$ cm$^{-2}$) (3) & 16.86 $\pm$ 0.58 & 8.63 $\pm$ 0.32 \\
    SFR (M$_{\odot}$ yr$^{-1}$) (4) & 43.5 $\pm$ 9.1 & 9.6 $\pm$ 1.3 \\
    sSFR (Gyr$^{-1}$) (5) & 2.01 & 14.6 \\
    \Ha/\Hbeta\ (6) & 4.68 $\pm$ 0.46 & 3.06 $\pm$ 0.16 \\
    $\rm E(B-V)$ (7) & 0.49 $\pm$ 0.10 & 0.07 $\pm$ 0.05 \\
    \NII/\Ha\ (8) & 0.12 $\pm$ 0.008 & 0.06 $\pm$ 0.008 \\
    \OIII/\Hbeta\ (9) & 4.09 $\pm$ 0.52 & 6.58 $\pm$ 0.30 \\
    \OIII/\OII\ (10) & 1.61 $\pm$ 0.15 & 4.48 $\pm$ 0.19 \\
    \enddata
    \tablenotetext{}{\textbf{Notes:} (1) Systemic redshift determined from nebular emission lines; (2) Stellar mass; (3) \Ha\ flux observed by Keck-MOSFIRE; (4) Star formation rate determined from the \Ha\ flux corrected for dust using the \Ha /\Hbeta\ ratio and the SMC extinction law; (5) Specific star formation rate; (6) Ratio of \Ha\ and \Hbeta\ spectral lines; (7) Reddening due to dust determined using measurements of the Balmer decrement and the SMC extinction law; (8) Ratio of \NII $\lambda$6583 and \Ha\ lines; (9) Ratio of \OIII\ $\lambda$5007 and \Hbeta\ lines; (10) Ratio of \OIII\ $\lambda\lambda$5007, 4959 and \OII\ $\lambda\lambda$3727, 3729 lines corrected for dust extinction}
\end{deluxetable} 

Additionally, we detect two previously unknown continuum sources, CS7 and CS10, in KCWI's field of view during our observations of BX710 and BX711. Both of these galaxies are located at a tentative spectroscopic redshift of $z=2.29$ and are denoted by labeled red circles in an \textit{HST}/ACS F814W image of the field, which is shown in the top panel of \ref{fig:im_lya}. We find that CS10 is a star-forming galaxy with a stellar mass of $M_{*} = 2.8 \times 10^{9} \ M_{\odot}$ located $28.7$ proper kpc south of BX710, and that CS7 is an $L_{*}$ star-forming galaxy with a stellar mass of $M_{*} = 1.9 \times 10^{10} \ M_{\odot}$ located just $10.4$ proper kpc east of BX711. The stellar masses we report for these galaxies are estimated from SED modeling assuming a redshift of $z=2.29$ for each galaxy. Despite CS7 and CS10's uncertain redshifts, the SED models fit their photometric measurements well. Nevertheless, due to the low S/N ratios of these galaxies' KCWI spectra, which causes their redshifts to be uncertain, we leave further analysis of these galaxies' spectra to future studies.
 
As an aside, we note that \Ha\ emission from CS7 and BX711 was detected in narrowband \Ha\ imaging from the Palomar 5.1m Hale telescope, which is presented in \citet{Bogosavljevic_phd_thesis}. Their regions of \Ha\ emission were originally attributed to a single source, BX711, but appear to be conjoined, with a surface brightness peak located in between the two galaxies. 

\subsection{Observations and Data Reduction}
\label{subsec:obsdata}

During the clear nights of 28 September 2019 and May 17 2020, we observed BX710 and BX711 using the Keck Cosmic Web Imager (KCWI), an optical integral field spectrograph on the Keck II Telescope on Mauna Kea \citep{Morrissey_2018}. BX710 ($\alpha_{\rm J2000}$ = 17:01:22.133, $\delta_{\rm J2000}$ = +64:12:19.30) and BX711 ($\alpha_{\rm J2000}$ = 17:01:21.288, $\delta_{\rm J2000}$ = +64:12:20.66) were observed simultaneously in 1200s intervals for a total exposure time of 5.1 hours. KCWI's BL grating and medium slicer were implemented during these observations, such that a $16.5 \arcsec \times 20.4 \arcsec$ area of sky centered around the galaxy pair and a wavelength range of $3530-5560$ \AA \ were simultaneously observed. 

Afterwards, the observational data were reduced into a 3D datacube via the KCWI Data Reduction Pipeline \footnote{https://github.com/kcwidev/kderp} following the procedure described in \citet{Chen_2021}. The resulting 3D KCWI datacube contains spatial dimensions of Right Ascension and Declination as well as a wavelength dimension. Additionally, the datacube's $16.5 \arcsec \times 20.4 \arcsec$ spatial dimensions are subdivided into individual spaxels, each of which spans a $0.3 \arcsec \times0.3 \arcsec$ region and a wavelength range of 1 \AA. KCWI's spectral resolution of 2.2 \AA \ at an observed wavelength of 4000 \AA \ ($R \approx 1800$) is well suited to analyzing the \lya\ emission throughout BX710 and BX711's \lya\ halos \citep{Chen_2021}.

 In order to determine whether BX710 and BX711 may be related to large scale structures in the HS1700+64 protocluster, we used $\mathcal{R}$-band observations of the field from the 4.2 m William Herschel Telescope. These observations were taken on the clear night of 24 May 2001 and covered a $15.4 \arcmin \times 15.4 \arcmin$ area of sky centered around the bright quasar HS1700+64. The resulting field image is described in \citet{Steidel_2004}.

\section{Spatial Measurements of the \lya\ Halos and Galaxy Filaments} \label{sec:sm}

The spatial extent of BX710 and BX711's \lya\ halos, as well as other spatial measurements of interest, can be measured using pseudo-narrowband images generated from the 3D KCWI datacube.

\subsection{Continuum Subtraction}
\label{subsec:continuumsubtraction}
In order to create a continuum-subtracted image of BX710 and BX711's \lya\ halos, we generated 1D optimally extracted spectra (i.e., down-the-barrel spectra) for each galaxy, from which we measured each galaxy's spectral continuum level. We produced these 1D spectra by spatially summing the KCWI datacube over isophotal apertures containing the galaxies' continuum emission over a wavelength range of $\sim980-1770$ \AA \ in the rest frame, as described in \citet{Erb_2023}. Circular apertures with the same area as the isophotal apertures would have radii of $\sim1.6\arcsec$ and $\sim1.3 \arcsec$ for BX710 and BX711, respectively. 
As in \citet{Chen_2021_phd_thesis}, we fit a BPASS v2.2 stellar population model \citep{Stanway_Eldridge_2018} with a metallicity of $Z=0.002$ and an age of $t=10^8$ yrs to these 1D spectra. We then scaled the resulting continuum fit to the individual spaxels and subtracted it from each spaxel of the 3D KCWI datacube within the isophotal apertures. 

\subsection{KCWI Images} \label{ssec:im}
\begin{figure*}[ht!]
    \centering
    \includegraphics[scale=0.5]{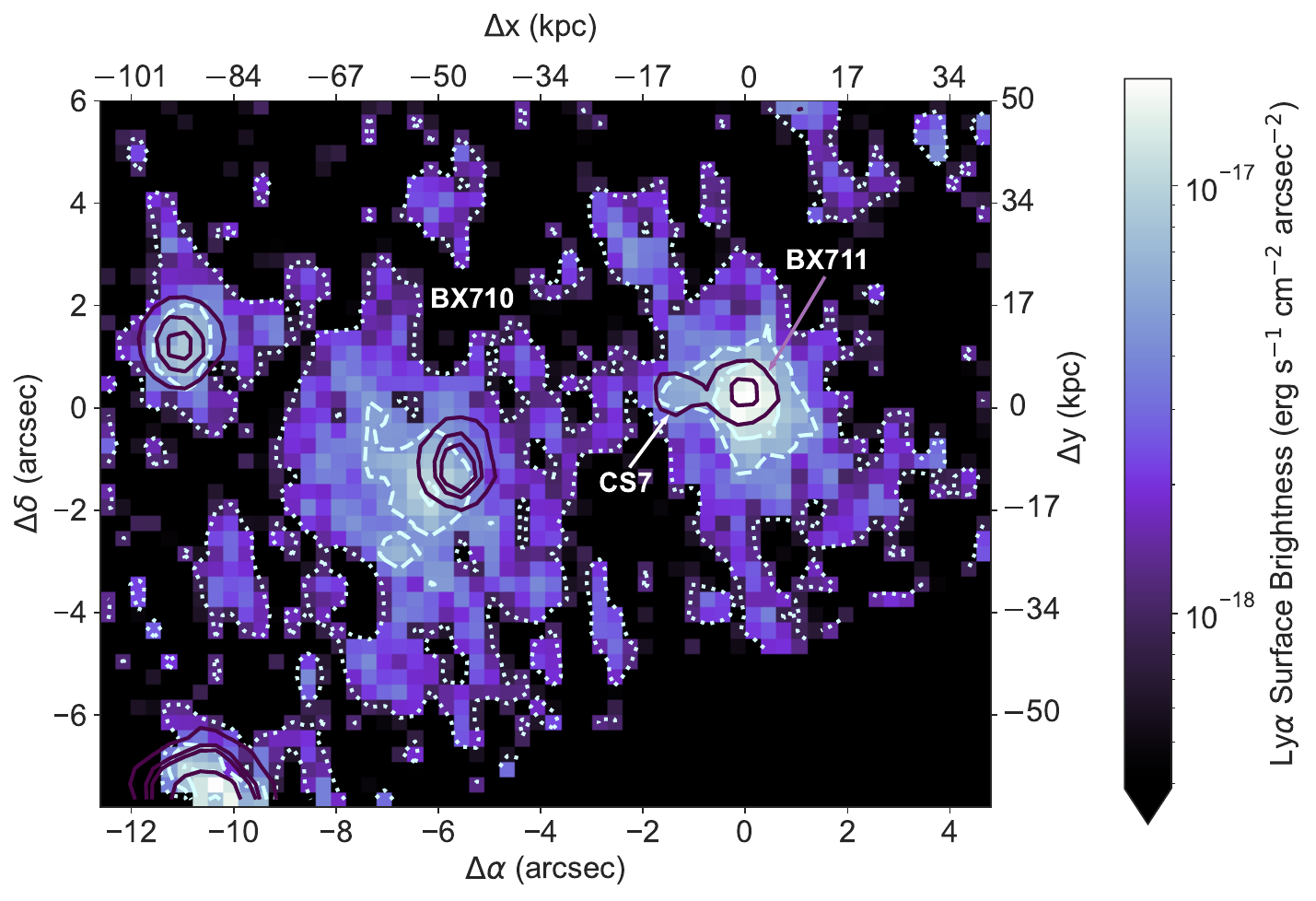}
    \centering
    \includegraphics[scale=0.63]{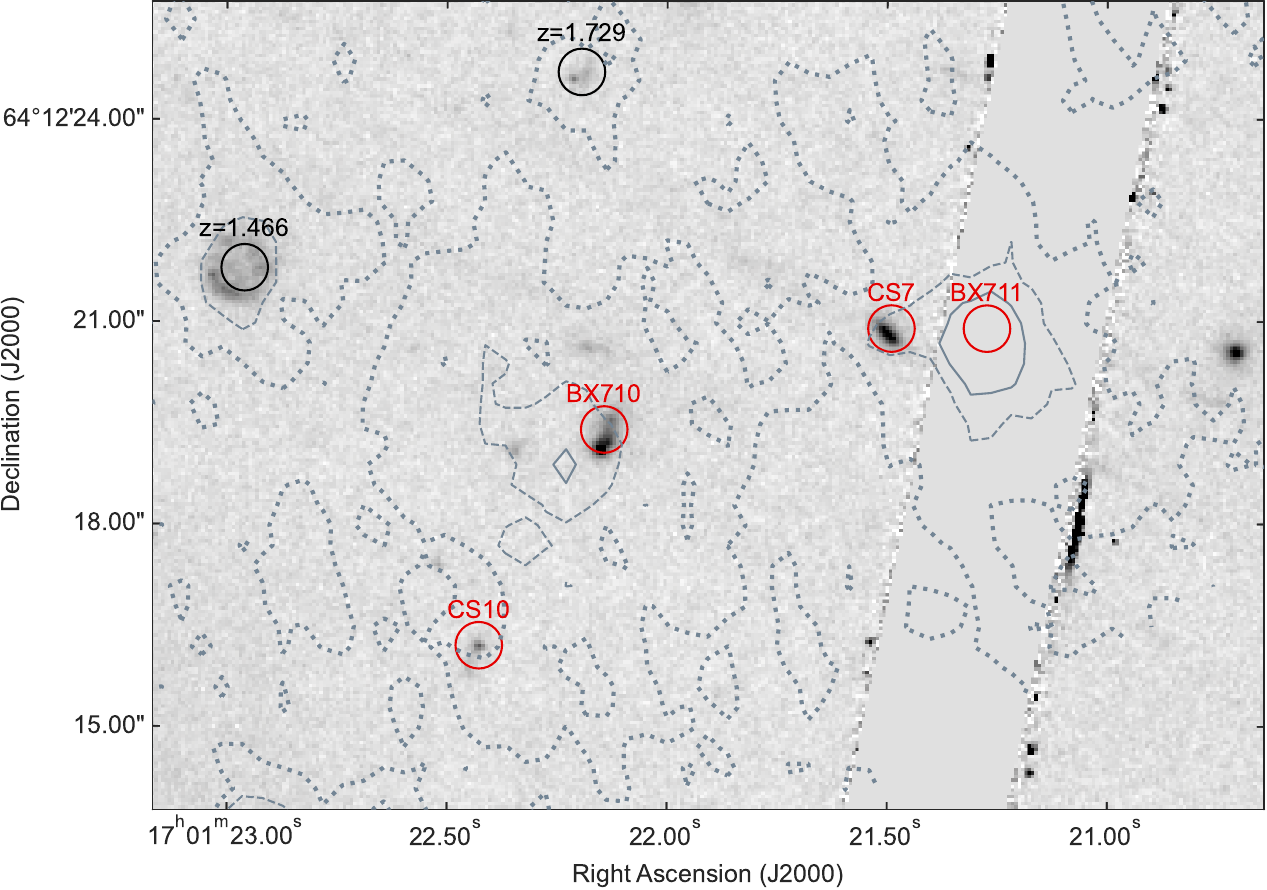}
    \caption{\textit{Top:} Pseudo-narrowband KCWI image of BX710 and BX711's \lya\ halos. UV continuum emission contours above a minimum flux threshold of $1.4 \times 10^{-16}$ erg \AA$^{-1}$ arcsec$^{-2}$ cm$^{-2}$ s$^{-1}$ are overlaid in deep violet. \lya\ emission contours with values of $[1, 5, 10] \times 10^{-18}$ erg arcsec$^{-2}$ cm$^{-2}$ s$^{-1}$ are delineated by ice blue dots, dashes, and dashes and dots, respectively. Other galaxies which lie at similar redshifts in this imaged region include CS10 and CS7, the latter of which lies within BX711's \lya\ halo and is visible as a protruding bulge on the eastern side of BX711's UV contours. Additionally included in this image is the continuum-selected galaxy BM583, an interloper at $z = 1.4664$ located northeast of BX710's \lya\ halo. All of the aforementioned galaxies are denoted by circles in the \textit{HST} image in the lower panel. \textit{Bottom:} \textit{HST}/ACS F814W image of BX710 and other targets in the field. BX711 is unfortunately not included among the targets imaged here, as it falls in the gap between detectors. The ice blue \lya\
    contours from the KCWI image in the upper panel are projected onto this image in dark blue. A more in-depth description of this image can be found in \citet{Peter_2007}.}
    \label{fig:im_lya}
\end{figure*}

 From the continuum-subtracted datacube, we generated a $16.8\arcsec \ \times 13.8\arcsec$ \ subcube containing BX710 and BX711's \lya\ halos using the Python package \textit{spectral-cube}. This \lya\ subcube spans a wavelength range of $3996 - 4018$ \AA \ in the observed frame ($\approx 1213-1220$ \AA \ in the rest frame), which was determined upon measuring the extent of the \lya\ profiles in the galaxy pair's 1D spectra. Subsequently, we integrated the continuum-subtracted \lya\ subcube over the wavelength axis, thereby creating a continuum-subtracted, pseudo-narrowband image of the \lya\ emission from these galaxies, which is displayed in Figure \ref{fig:im_lya}. Revealed in this image are BX710 and BX711's \lya\ halos, which are comprised of extended, elongated regions of \lya\ surface brightness surrounding each galaxy's ISM.  BX710's \lya\ halo is $\sim 50$ pkpc wide and $\sim 80$ pkpc long, while BX711's \lya\ halo is $\sim 40$ pkpc wide and $\sim 70$ pkpc long. 

    Similarly, from the pre-continuum-subtracted 3D datacube we extracted a subcube of the UV emission which covers the same region as the \lya\ subcube and a wavelength range of $4040 - 5450$ \AA \ in the observed frame (i.e., essentially $G$-band; $\approx 1226-1654$ \AA \ in the rest frame). We then created a UV image by taking the mean continuum value within this wavelength range for each spaxel. The royal blue UV contours in the upper panel of Figure \ref{fig:im_lya} denote where the galaxies' star forming regions are located. Comparison of the \lya\ and UV contours in Figure \ref{fig:im_lya} reveals a marked spatial offset between the spaxels with the greatest \lya\ and UV surface brightness for BX710 of $\Delta d^{\rm BX710}_{ \rm UV-Ly\alpha} = 7.1$ proper kpc, while for BX711 the brightest UV and \lya\ spaxels are aligned within one spaxel. This unusually large offset between BX710's brightest UV and \lya\ spaxels could result from gaps in the HI gas in the CGM \citep{Leclercq_2017, Ribeiro_2020, Claeyssens_2022}. Additionally, we determined the distance between BX710 and BX711's brightest \lya\ spaxels to be $d_{\rm Ly\alpha} = 56$ proper kpc.

\subsection{ Galaxy Filaments in the HS1700+64 Protocluster} \label{ssec:fil}
\begin{figure*}[ht] 
\includegraphics[scale = 0.75]{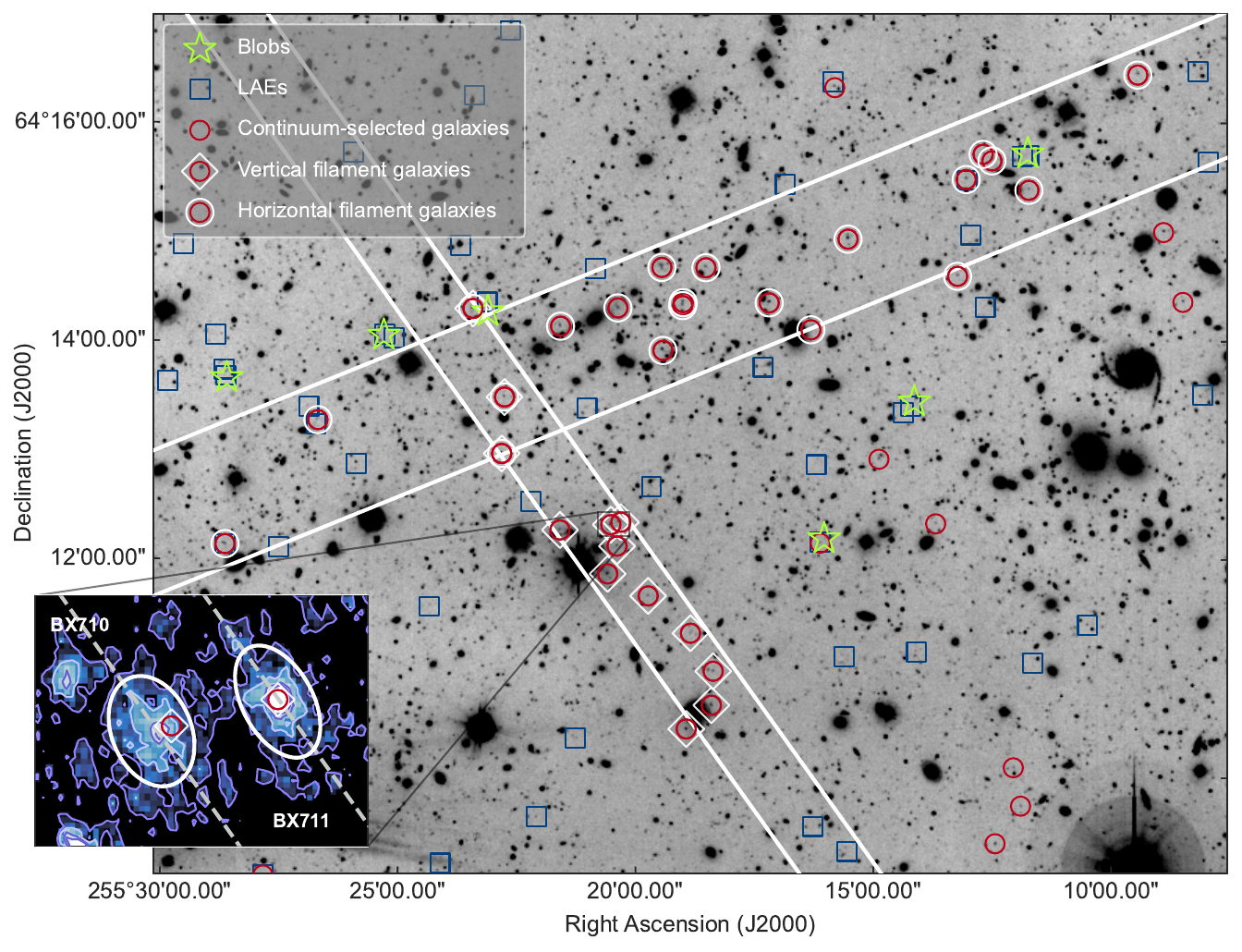}
\caption{The HS1700+643 protocluster field containing Q1700-BX710, Q1700-BX711, and a multitude of other UV continuum-selected galaxies that have been spectroscopically confirmed to be protocluster members (all in dark red). The bounds of the negatively-sloped and positively-sloped filaments are indicated by white diagonal lines. UV-continuum selected galaxies residing in the negatively-sloped filament are represented by dark red circles surrounded by white diamonds, whereas UV-continuum selected galaxies residing in the positively-sloped filament are represented by dark red circles surrounded by white concentric circles. \textit{Zoomed-in inset plot}: Elliptical isophote fits (white) to BX710 and BX711's \lya\ halos (on left and right, respectively). Here, the negatively-sloped filament is projected onto the centers of BX710 and BX711's \lya\ halos as two parallel, silver dashed lines intersecting their elliptical isophotes in order to easily compare the angles of the filament and the elliptical isophotes. The dashed lines do not represent the center of the filament or denote its width.}
\label{fig:lss}
\end{figure*}

The $\mathcal{R}$-band image of the HS1700+64 field in Figure \ref{fig:lss} shows that a multitude of galaxies are scattered throughout this overdense environment, including both UV-continuum selected galaxies such as Q1700-BX710 and Q1700-BX711 and \lya\ emitters (LAEs) selected via narrowband imaging. The color selection criteria for these UV continuum-selected and \lya\-selected galaxies are described in \citet{Steidel_2004} and \citet{Erb_2014}, respectively. 

As seen in the protocluster field image, BX710 and BX711 appear to be part of a quasilinear filament consisting of numerous continuum-selected galaxies. Because of this, we performed least squares linear regression on the positions of continuum-selected galaxies which seemed to reside within or near this group of galaxies. During the fitting process, our first priority was selecting a combination of galaxies to be part of this filament such that the number of galaxies in the filament was maximized, and our second priority was maximizing the correlation coefficient. We found that this filament is indeed approximately linear ($R^2 = 0.895$), contains 13 continuum-selected galaxies, and is oriented at $31.2 \pm 3.6 \degree$ east of north. We defined the width of the filament to be twice the standard deviation of the perpendicular distances from the galaxies to the filament's regression line. The resulting width is 1.48 comoving Mpc. In Figure \ref{fig:lss}, this filament is displayed as two diagonal white lines which denote the filament's width.

Using the same method, we define another galaxy filament in this field including 21 UV continuum-selected galaxies, that intersects the negatively-sloped filament containing BX710 and BX711 at an angle. We find that this filament is broader ($R^2 = 0.83$), is oriented at $10.1 \pm 1.1 \degree$ west of north, and has a width of 2.18 comoving Mpc. Intriguingly, this filament appears to be roughly aligned with four out of six total "\lya\ blobs" (LABs) reported by \citet{Erb_2011} in the HS1700+64 protocluster, which are arranged linearly.

Note that we excluded compact LAEs located in the HS1700+64 protocluster from consideration as members of the galaxy filaments because they appear to be fairly randomly distributed throughout the protocluster. Two-thirds of all spectroscopically-identified UV-continuum selected galaxies in this protocluster are contained within these two filaments. However, we warn that not all of the UV-continuum selected galaxies in this protocluster have spectroscopically confirmed redshifts. Additionally, many of those that do were identified from NIR imaging by the Palomar 5.1m Hale telescope in a $\sim 9 \arcmin \times 9 \arcmin$ region within the protocluster field \citet{Erb_2006} and have subsequently been targeted the most heavily by followup spectroscopic observations.

Several galaxy filaments similar to the two which we have identified in the HS1700+64 protocluster have been reported in the last couple of decades. \citet{Matsuda_2005} reported the existence of three filamentary structures consisting of 56 LAEs as well as multiple LBGs and the \lya\ blobs LAB1 and LAB2 in the SSA22 protocluster. Aided by the availability of spectroscopic redshifts for the LAEs contained in these filaments, \citet{Matsuda_2005} overlaid a contour for the total volume density of all 56 LAEs on the protocluster image, revealing three intersecting filaments within the SSA22 protocluster. Other papers, including \citet{Harikane_2019}, \citet{Umehata2019}, \citet{Daddi_2021}, and \citet{Bacon_2021}, have uncovered filaments of \lya\ emission in overdense environments containing populations such as LAEs, LBGs, \lya\ blobs, submillimeter galaxies (SMGs), dusty star-forming galaxies (DSFGs), and active galactic nuclei (AGN). All of these papers define filamentary structures using a \lya\ surface brightness threshold.

\subsection{\lya\ Halo Alignment with Galaxy Filament and Spatial Extent}
Whereas most \lya\ halos are fairly circular \citep[e.g.,][]{Matsuda_2012, Wistotzki_2016, Leclercq_2017}, it is apparent from Figure \ref{fig:im_lya} that BX710 and BX711's \lya\ halos are elongated, and from the projections of the negatively-sloped filament indicated by dashed white lines in the Figure \ref{fig:lss} inset that the direction of elongation coincides with the position angle of the filament. Such an alignment may signal that narrow streams of low-metallicity gas flowing among the galaxies in the filament are accreting onto BX710 and BX711 \citep{Umehata2019}. 

For this reason, we fit elliptical isophotes to their \lya\ halos using the continuum-subtracted \lya\ image (see Figure \ref{fig:im_lya}) and the Python package \textit{photutils} \citep{Bradley_2022}, as shown in the inset to Figure \ref{fig:lss}. The resulting elliptical isophotes for each galaxy's halo are displayed in white in the inset image of Figure \ref{fig:lss}. 

From these elliptical isophotes, we ascertained quantities such as the estimated size and position angles of these galaxies' \lya\ halos. We found that BX710's \lya\ halo is oriented at $22.9 \pm 2.3 \degree$ east of north, has a semi-major axis of $25.1$ proper kpc, and has a semi-minor axis of $17.3$ proper kpc. Likewise, BX711's \lya\ halo is oriented at $29.0 \pm 2.4 \degree$ east of north, has a semi-major axis of $26.6$ proper kpc, and has a semi-minor axis of $15.3$ proper kpc. Both are approximately aligned with the galaxy filament oriented at $31.2 \pm 3.6 \degree$ east of north within their uncertainties.

\section{\lya\ and UV Spectra} \label{sec:spectra}
    
    \subsection{Voronoi Binning} \label{ssec:binning}
\begin{figure}[ht!]  
    \centerline{\epsfig{angle=00,width=\hsize,file=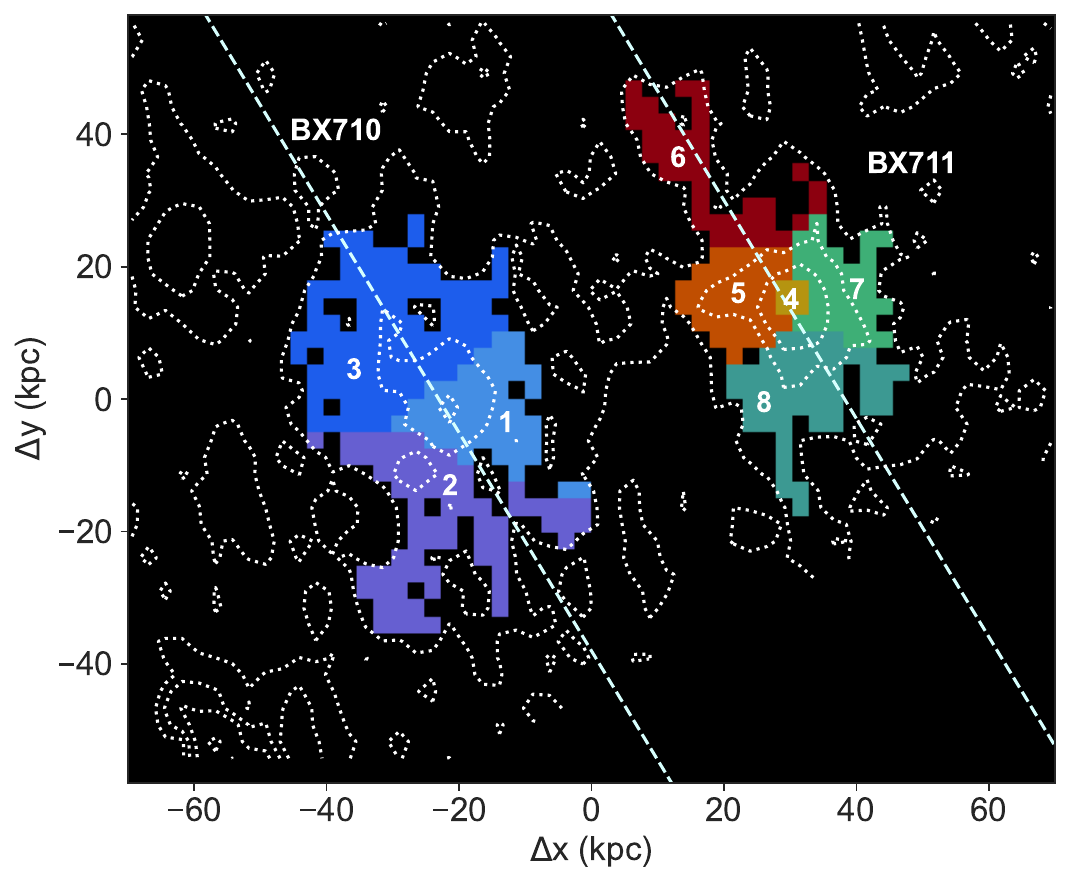}}
    \caption{Image of the Voronoi- binned regions in BX710 and BX711's \lya\ halos. Dotted white lines delineate contours of \lya\ emission with the same levels as in Figure \ref{fig:im_lya}.}
    \label{fig:bin_im}
    \end{figure}

    Most of the individual spaxels comprising BX710 and BX711's \lya\ halos have low S/N ratios, so we implemented a Voronoi binning procedure as described by \citet{Cappellari_2003} to subdivide the \lya\ halos into several regions with reasonably high S/N \lya\ spectra using the Python package \textit{vorbin}. We performed Voronoi binning on contiguous spaxels with S/N $>$ 1 in the \lya\ halos displayed in the pseudo-narrowband, continuum-subtracted \lya\ image (Figure \ref{fig:im_lya}). This process resulted in 3 smaller Voronoi binned regions for BX710, covering an area of $\sim 19$ arcsec$^2$ with \lya\ spectra with S/N ranging from $\sim 14-18$. For BX711, 5 smaller Voronoi binned regions were produced with a total area of $\sim 14$ arcsec$^2$ and \lya\ S/N ranging from $\sim 10-18$. These Voronoi bins are displayed in Figure \ref{fig:bin_im}.

    \subsection{Spatially Resolved \lya\ Spectra}
\begin{figure*}[ht!]
    \centering
    \includegraphics[scale = 0.75]{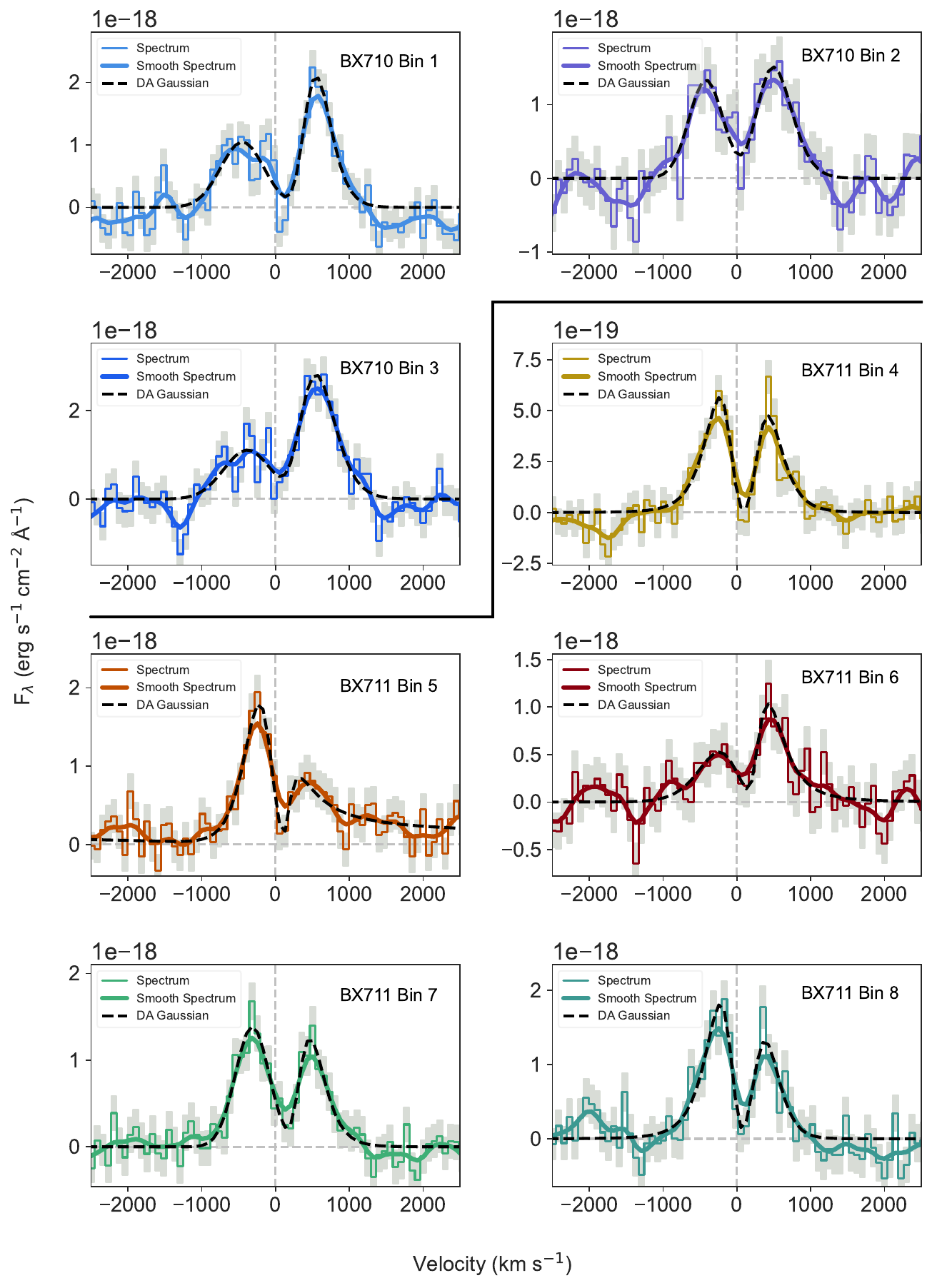}
    \caption{Velocity space \lya\ spectra of the Voronoi binned regions in BX710 and BX711's \lya\ halos. The numerical label for each spectrum corresponds to a region with the same label in Figure \ref{fig:prps}, and the color of the spectrum likewise corresponds to the color of that region. Solid curves with the same color as the spectra represent the spectra's appearances after being smoothed by a 1D Gaussian kernel with a size of 13. The double asymmetric Gaussian fits to these spectra are displayed as black dashed curves, and the dashed silver vertical line indicates the systemic velocity.}
    \label{fig:voronoi-spectra}
    \end{figure*}

    We extracted \lya\ spectra from each of the Voronoi binned regions by spatially summing the continuum-subtracted datacube over all of the spaxels in each bin. The resulting continuum-subtracted \lya\ spectra, all of which have double-peaked profiles, are displayed in Figure \ref{fig:voronoi-spectra}.

\begin{figure*}[ht!] \label{fig:prps}
        \centering
        \includegraphics[scale = 0.4]{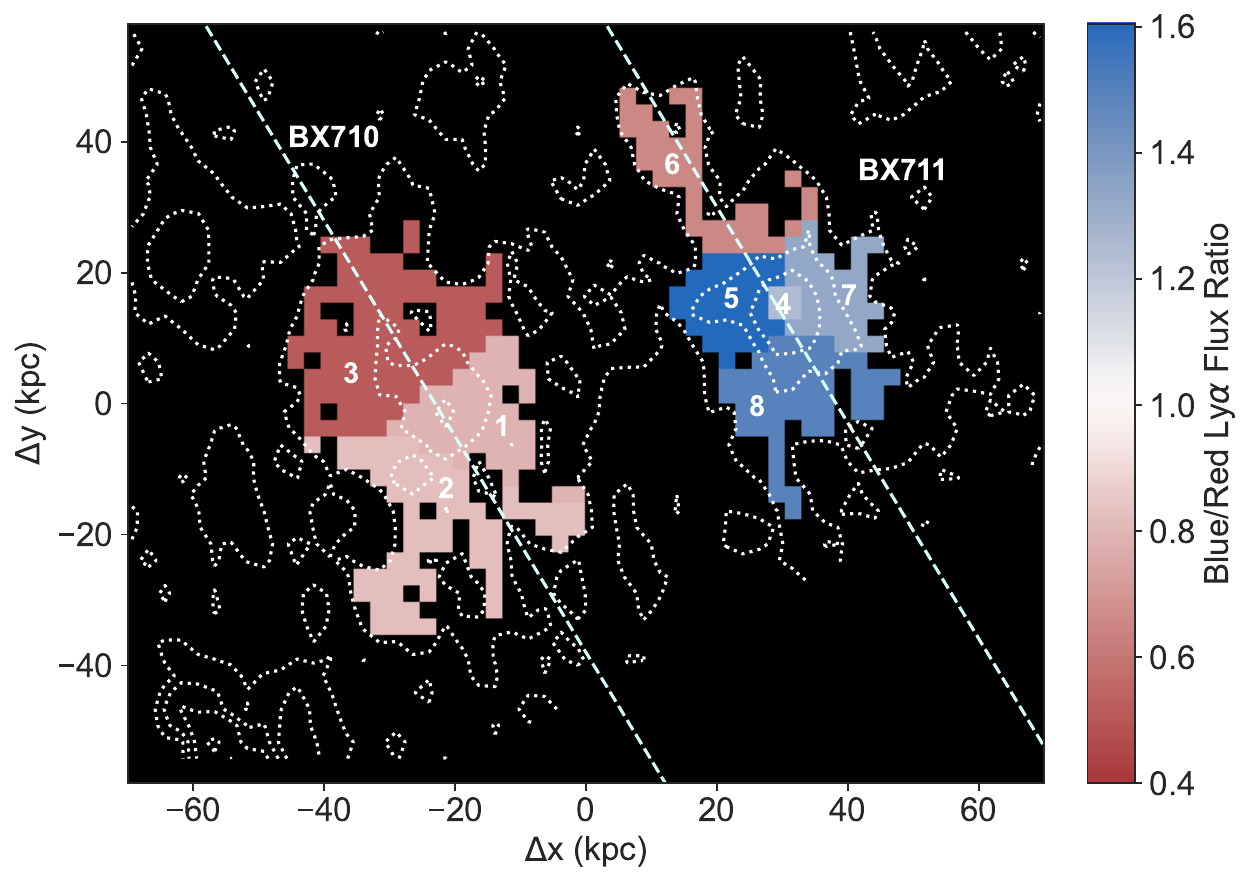}
        \includegraphics[scale = 0.4]{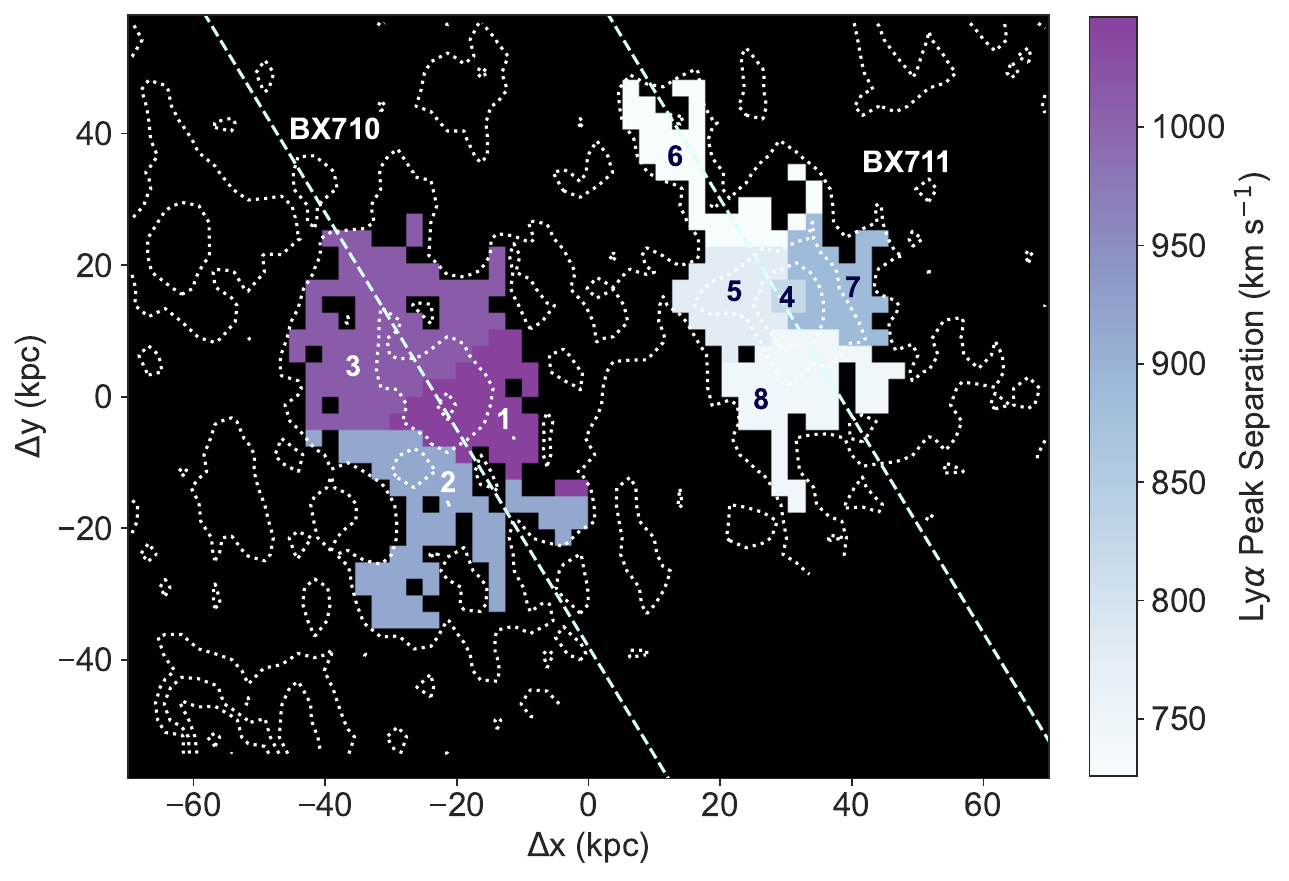}
    \caption{\textit{Left}: Blue-to-red peak ratio image of BX710 and BX711's \lya\ halos. \textit{Right}: Peak separation image of BX710 and BX711's \lya\ halos. Dotted white lines in both panels show contours of \lya\ emission with the same levels as in Figure \ref{fig:im_lya}.}
    \end{figure*}

    For each resulting \lya\ spectrum, we defined the lower velocity boundary of the blueshifted peak and upper velocity boundary of the redshifted peak to be where the S/N per pixel drops below 2 on their blueward and redward sides, respectively. The upper velocity boundary of the blueshifted peak and the lower velocity boundary of the redshifted peak, on the other hand, are both located at the minimum between the two peaks. Using these velocity boundaries, we then integrated the flux from each peak and calculated the blue-to-red peak flux ratio for each \lya\ profile from the Voronoi binned regions. Our resulting measurements of the spatially varying blue-to-red peak flux ratio throughout BX710 and BX711's \lya\ halos are displayed in the left panel of Figure \ref{fig:prps} and listed in Table \ref{tab:prps_tab}. 
    
    Additionally, for each \lya\ spectrum we calculated flux-weighted centroids corresponding to each peak using the velocity boundaries described above. By subtracting the red component's centroid from the blue component's centroid, we measured the peak separation for each bin in velocity space. The spatially varying peak separation throughout BX710 and BX711's \lya\ halos is shown in the right panel of Figure \ref{fig:prps} and listed in Table \ref{tab:prps_tab}.
    
    As shown in the peak separation and peak ratio maps, BX710 is primarily characterized by dominant red peaks and large peak separations ranging from $\approx 900-1050$ kms$^{-1}$, while BX711 is primarily characterized by dominant blue peaks and narrower peak separations ranging from $\approx 725-900$ kms$^{-1}$. Some of BX711's regions, strikingly, have blue-to-red peak flux ratios as high as around 1.6. A notable exception to this pattern is a single red dominated region in the northernmost part of the halo. 
    
    \begin{deluxetable}{c c c}[h!]
    \renewcommand{\arraystretch}{1.25}
    \tablecaption{\lya\ Spectral Properties}
    \label{tab:prps_tab}
    \tablehead{\multicolumn{1}{c}{Bin}&\multicolumn{1}{c}{Blue-to-red Peak Flux Ratio}&\multicolumn{1}{c}{Peak Separation (km s$^{-1}$)}}
    \startdata
    BX710 & & \\
    \hline
    1 & 0.79 $\pm$ 0.09 & 1046 $\pm$ 48\\
    2 & 0.82 $\pm$ 0.12 & 915 $\pm$ 57\\
    3 & 0.51 $\pm$ 0.07 & 1012 $\pm$ 62\\
    \hline
    BX711 & & \\
    \hline
    4 & 1.24 $\pm$ 0.15 & 822 $\pm$ 44\\
    5 & 1.60 $\pm$ 0.21 & 777 $\pm$ 47\\
    6 & 0.65 $\pm$ 0.14 & 726 $\pm$ 62\\
    7 & 1.32 $\pm$ 0.16 & 885 $\pm$ 55\\ 
    8 & 1.49 $\pm$ 0.20 & 742 $\pm$ 43\\ 
    \enddata
    \tablenotetext{}{\textbf{Notes:} Blue-to-red \lya\ peak flux ratios and peak separations for the Voronoi bins in BX710 and BX711's \lya\ halos, which are denoted by corresponding numbered labels in Figure \ref{fig:prps}. Voronoi bins 1-3 are located in BX710's \lya\ halo, and bins 4-8 are located in BX711's \lya\ halo.}
    \end{deluxetable} 
    
    \subsubsection{Double Asymmetric Gaussian Fits}
    Previous studies of galaxies with double-peaked \lya\ emission have found that a double asymmetric Gaussian describes these \lya\ profiles well \citep{Trainor_2015, Erb_2023, Mukherjee_2023}. For this reason, we fit the \lya\ profiles in each binned region's spectrum with a double asymmetric Gaussian model. This model is defined by Equation 1 in \citet{Erb_2023}, as:
    \begin{multline}
        f(v) = A_{\rm blue}\exp\biggl(\frac{-(v - v_{0, \rm blue})^2}{2\sigma_{\rm blue}^2}\biggl) \\
        + A_{\rm red}\exp\biggl(\frac{-(v - v_{0, \rm red})^2}{2\sigma_{\rm red}^2}\biggl)
    \end{multline}
    where $A_{\rm blue}$ and A$_{\rm red}$ are the respective amplitudes of the blueshifted and redshifted \lya\ peaks and $v_{0, \rm blue}$ and v$_{0, \rm red}$ correspond to the velocity at the center of each of these peaks. Furthermore, the width $\sigma$ of each peak is given by the equation $\sigma = a(v - v_0) + d$, where $a$ is the asymmetry parameter, which has a range of $-1 < a < 0$ for the blueshifted peak and a range of $0 < a < 1$ for the redshifted peak, and $d$ is a parameter for each peak's width \citep{Leclercq_2020, Erb_2023}.
    
    Best-fit model profiles to the \lya\ spectra are shown in Figure \ref{fig:voronoi-spectra}. We find that the \lya\ profiles throughout BX710 and BX711's \lya\ halos are described well by these fits.
    
    \subsection{UV Features}\label{sec:uvfeatures}
    Whereas \lya\ potentially traces the kinematics of \HI\ gas, other UV features reflect the kinematics of metal-enriched gas in the foreground of the galaxy that is not as strongly affected by resonant scattering, and can thereby help paint a more complete picture of the gas flows in the CGM than by studying \lya\ alone. For this reason, we measured all of BX710 and BX711's UV features from their 1D optimally extracted UV spectra described in Section \ref{subsec:continuumsubtraction}, as these galaxies' star-forming regions do not significantly overlap most of the Voronoi-binned regions.
    
    Within these 1D spectra, we detected multiple low-ionization and high-ionization absorption features commonly found among LBGs \citep{2003ApJ...588...65S, Jones_2012}. Specifically, the low-ionization features we identified include \SiII\ $\lambda 1260$, \OI\ $\lambda 1302$, \SiII\ $\lambda 1304$, \CII\ $\lambda 1334$, \SiII\ $\lambda 1526$, \FeII\ $\lambda 1608$, and \AlII\ $\lambda 1670$, while the high-ionization absorption features include \SiIV\ $\lambda 1393$, \SiIV\ $\lambda 1402$, and \CIV\ $\lambda\lambda 1548, 1550$.  Additionally, we identified several fine-structure emission lines, including the less commonly measured \SiIIfs\ $\lambda 1197$ line, \SiIIfs\ $\lambda\lambda 1264, 1265$, \SiIIfs\ $\lambda 1309$, and \SiIIfs\ $\lambda 1533$. 
    
    We measured the equivalent width and velocity offset for each UV feature except \lya\ directly from the optimally extracted, continuum-normalized spectra. Here we define the velocity offset $\Delta v$ for each UV feature as the difference in velocity between a feature's flux-weighted centroid and the systemic redshift of the galaxy in question, such that $\Delta v = v_{centroid} - v_{sys}$. The equivalent widths and velocity offsets for UV features with S/N $\gtrsim 3$ are listed in Table \ref{tab:EW}. 
    
    \begin{deluxetable*}{c c c c c c }[t!]
    \tablecaption{Q1700-BX710 and Q1700-BX711 UV Measurements}
    \label{tab:EW}
        \tablehead{\colhead{UV Feature} & \colhead{Rest Wavelength} & \colhead{EW$_{\rm BX710}$} & \colhead{$\Delta v_{\rm BX710}$} & \colhead{EW$_{\rm BX711}$} & \colhead{$\Delta v_{\rm BX711}$} \\
        \colhead{ } & \colhead{(\AA)} & \colhead{(\AA)} & \colhead{(km s$^{-1}$)} & \colhead{(\AA)} & \colhead{(km s$^{-1}$)} \\
        \colhead{(1)} & \colhead{(2)} & \colhead{(3)} & \colhead{(4)} & \colhead{(5)} & \colhead{(6)}}   
    \startdata
     & & Low-ionization Interstellar (LIS) Features & & & \\
     \hline
    \SiII\ $\lambda 1260$ & $1260.42$ & $-2.05 \pm 0.25$ & $-22 \pm 27$ & $>-0.93$ & -- \\ 
    \OI\ $\lambda 1302$ & $1302.17$ & $-1.97 \pm 0.16$ & -- & -- & -- \\
    \SiII $\lambda 1304$ & $1304.37$ & $-1.36 \pm 0.17$ & -- & -- & -- \\
    \CII\ $\lambda 1334$ & $1334.53$ & $-2.08 \pm 0.19$ & $109 \pm 17$ & $-1.23 \pm 0.26$ & $11 \pm 34$ \\
    \SiII\ $\lambda 1526$ & $1526.71$ & $-2.14 \pm 0.20$ & $35 \pm 24$ & $-1.05 \pm 0.24$ & $51 \pm 37$ \\
    \FeII\ $\lambda 1608$ & $1608.45$ & $-1.92 \pm 0.24$ & $245 \pm 35$ & -- & -- \\
    \AlII\ $\lambda 1670$ & $1670.79$ & $-1.47 \pm 0.21$ & $190 \pm 25$ & -- & -- \\
    \hline
    & & High-ionization Interstellar (HIS) Features & & & \\
    \hline
    \SiIV\ $\lambda 1393$ & $1393.76$ & $-1.29 \pm 0.21$ & $-55 \pm 33$ & $-1.24 \pm 0.27$ & $-74 \pm 37$ \\
    \SiIV\ $\lambda 1402$ & $1402.77$ & $-0.72 \pm 0.16$ & $32 \pm 29$ & $-0.82 \pm 0.22$ & $24 \pm 31$ \\
    \CIV\ $\lambda 1548$ & $1548.20$ & $-3.70 \pm 0.29$\tablenotemark{a} & $-691 \pm 43$\tablenotemark{b} & $-1.60 \pm 0.31$ & $-140 \pm 39$ \\
    \CIV\ $\lambda 1550$ & $1550.78$ & $-3.70 \pm 0.29$\tablenotemark{a} & $-691 \pm 43$\tablenotemark{b} & $-0.86 \pm 0.23$ & $-51 \pm 33$ \\
    \hline 
    & & Emission Features & & & \\
    \hline
    \SiIIfs\ $\lambda 1197$ & $1197.39$ & $1.51 \pm 0.26$ & $-68 \pm 23$ & -- & -- \\
    \lya\ & $1215.67$ & Emission: $11.69 \pm 1.43$ \ Total: $\sim -15$ & -- & $19.33 \pm 0.96$ & -- \\
    \SiIIfs\ $\lambda\lambda 1264, 1265$ & $1264.74$ & $1.30 \pm 0.24$ & $129 \pm 30$ & $1.37 \pm 0.35$ & $92 \pm 37$ \\
    \SiIIfs $\lambda 1309$ & $1309.28$ & $1.49 \pm 0.23$ & $-75 \pm 48$ & -- & -- \\
    \SiIIfs\ $\lambda 1533$ & $1533.43$ & $0.44 \pm 0.12$ & $51 \pm 38$ & -- & -- \\
    \enddata
            \tablenotetext{}{(1) UV features; (2) Rest wavelength for each UV feature; (3) Equivalent widths for BX710's UV features; (4) Velocity offsets for BX710's UV features; (5) Equivalent widths for BX711's UV features; (6) Velocity offset for BX711's UV features}
            \tablenotetext{a}{EW measurement of the blended \CIV $\lambda\lambda$ 1548, 1550 feature}
            \tablenotetext{b}{Velocity offset measurement of the blended \CIV $\lambda\lambda$1548, 1550 feature calculated with respect to the average rest wavelength of these features}
            
    \end{deluxetable*} 
    In BX710's 1D spectrum, the low-ionization \OI\ $\lambda$1302 and \SiII\ $\lambda$1304 features, which are often blended in LBG spectra, appear to be partially blended \citep{2003ApJ...588...65S, Jones_2012}. Specifically, taking KCWI's spectral resolution into consideration, these features have minima that are clearly distinct, while the redward wing of \OI\ $\lambda$1302 and the blueward wing of \SiII\ $\lambda$1304 are blended. As a result, we estimated their equivalent widths separately but excluded their measured velocity offsets from Table \ref{tab:EW} since they are not measured independently. 
    
    For BX710, the mean equivalent widths for its low ionization and high ionization features are $EW_{\rm 710, LIS} = -1.93 \pm 0.10$ \AA\ and $EW_{\rm 710, HIS} = -1.01 \pm 0.13$ \AA, respectively, with mean velocity offsets corresponding to $\Delta v_{\rm 710, LIS} = 111 \pm 12$ km s$^{-1}$ and $\Delta v_{\rm 710, HIS} = -11\pm 22$ km s$^{-1}$. We excluded \CIV$\lambda\lambda 1548, 1550$ from these calculations due to a broad stellar absorption component in its observed absorption profile. Likewise, the mean equivalent widths for BX711's low ionization and high ionization features are given by $EW_{\rm 711, LIS} = -1.14 \pm 0.18$ \AA\ and $EW_{\rm 711, HIS} = -1.13 \pm 0.13$ \AA, and its mean velocity offsets for these features are given by $\Delta v_{\rm 711, LIS} = 31 \pm 25$ km s$^{-1}$ and $\Delta v_{\rm 711, HIS} = -60 \pm 18$ km s$^{-1}$. The errors on the mean EWs and mean velocity offsets are determined from error propagation of the errors on the individual EWs and velocity offsets involved, whose uncertainties in turn stem from the 1D spectra. 

    In comparison to mean velocity offsets for low-ionization lines reported in the literature for similar galaxies at z $\gtrsim$ 2 (e.g., $\Delta v_{\rm LIS} = -150 \pm 60$ km s$^{-1}$ in \citet{2003ApJ...588...65S}, $\Delta v_{\rm LIS} = -190$ km s$^{-1}$ in \citet{Jones_2012}, and $\Delta v_{\rm LIS} = -165 \pm 125$ km s$^{-1}$ in \citet{Steidel_2014}), BX710 and BX711's low-ionization velocity offsets are substantially redshifted in the case of BX710 and slightly redshifted in the case of BX711, suggesting gas inflow. We discuss this possibility in Section \ref{sec:dis} below.
    
    As for \lya, absorption from the IGM lowered the continuum level significantly on the blueward side of the \lya\ profiles for both galaxies, so we measured the continuum levels on both sides of the \lya\ profile separately and from the 1D optimally extracted spectra rather than the normalized 1D spectra. In the case of BX711, whose \lya\ profile consists solely of emission, we used the median of the blueward and redward continuum windows from \citet{Kornei_2010}, which are given by 1120-1180 \AA \ and 1225-1255 \AA, respectively, to determine the continuum levels blueward and redward of the \lya\ emission profile. We then calculated the equivalent width for each peak separately using the blueward continuum level for the blueshifted peak and the redward continuum level for the redshifted peak. Finally, we added the equivalent widths from both peaks together to produce a total \lya\ emission equivalent width of $19.3 \pm 0.43$ \AA\ for BX711, which is recorded in Table \ref{tab:EW}. 
    The blue-to-red peak flux ratio for this \lya\ profile is 1.1, which is slightly lower than BX711's central Voronoi-binned regions, as shown in the left panel of \ref{fig:prps}.
    
    In the case of BX710, we found that its \lya\ profile appears to consist of emission superimposed on broad absorption, so we measured separate continuum levels for the emission profile and the total \lya\ profile. For BX710's \lya\ emission profile, we defined narrow blueward and redward continuum windows right next to the line, each with a width of 10 \AA, and then, as with BX711, took the median of each window to be the blueward and redward continuum level. Using these continuum levels, we measured the equivalent width for the \lya\ emission in the same manner as described above for BX711. For BX710's total \lya\ profile, we established windows with widths of 15 \AA \ on either side of the approximate extent of the broad \lya\ absorption. These windows did not include any other absorption or emission lines. By taking the continuum level surrounding the entire \lya\ profile to be the median of these two windows combined, we measured a rough estimate for the equivalent width of BX711's total \lya\ profile. The equivalent widths for BX710's \lya\ emission profile and total \lya\ profile, which are given by $11.7\pm0.49$ \AA\ and $\sim-15$ \AA, respectively, are listed in Table \ref{tab:EW}.

    Overall, from our analysis of BX710 and BX711's UV features we have learned that: 1) slightly blue-dominated \lya\ emission is present in BX711's star-forming regions as well as its CGM, suggesting that BX711 may be in the process of forming new stars from accreting gas, and 2) that all of BX710 and BX711's low-ionization metal absorption features are either redshifted or consistent with the systemic velocity of the galaxy, which, as previously mentioned, is unusual among $z\sim2-3$ galaxies and likewise indicates inflowing gas in agreement with \lya.
    
\section{Modeling \lya\ Emission And Low-ionization UV Absorption Lines} 
\label{sec:modeling}

In this section, we employ a set of radiative transfer (RT) models and semi-analytic models to simulate the spatially-resolved \lya\ emission profiles and down-the-barrel UV absorption line profiles of BX710 and BX711. 

\subsection{Spatially-resolved Ly$\alpha$ emission} 
\label{sec:lya_modeling}
We model the spatially-resolved \lya\ profiles of BX710 and BX711 with a grid of multiphase, clumpy RT models in a similar way to \citet{Erb_2023}, using the the 3D \lya\ Monte Carlo RT code \texttt{tlac} \citep{Gronke14}. The models are designed to resemble the structure of the ISM / CGM, and each model consists of two phases of gas: cool ($\sim 10^4$\,K) \HI\ clumps and a hot ($\sim 10^6$\,K), highly ionized interclump medium. In each model, Monte Carlo RT simulations are performed using 10$^4$ \lya\ photon packages, which emanate from the center of a spherical region. After interacting with the \HI\ content in both the cool and hot gas components, all \lya\ photons eventually escape\footnote{We do not consider the destruction of \lya\ photons due to dust extinction in this work.} from the simulation region. The \lya\ photons that escape from different impact parameters can be separated into several spatial bins and constitute spatially-resolved model spectra, which can be further used to compare with the corresponding observed spatially-resolved \lya\ profiles.

Our observations of BX710 show that the \lya\ profiles in the central region are red-peak-dominated and become less so in the outer regions. For BX711, however, the double peaks appear symmetric in the central region (although the red peak contains less flux), and become blue-peak-dominated in the outer regions to different degrees along different directions. We hereby assume that the radial motion of the cool clumps continuously transitions from an outflow to an inflow. Specifically, it is assumed that the clump radial motion reaches its maximum outflow velocity $+v_1$ at a certain radius $r_{\rm min}$, and gradually transitions (e.g.\ in the form of a power law) to a maximum inflow velocity $-v_2$ at radius $r_{\rm max}$, and its velocity profile can be written as:
\begin{equation}
v_{\rm cl}(r) = v_1 - (v_1 + v_2)\left(\frac{r - r_{\rm min}}{r_{\rm max} - r_{\rm min}}\right)^\alpha
\label{eq:radial_v}
\end{equation}
where $\alpha$ is the power-law index that determines the exact shape of the velocity profile. Note that such a radial velocity profile is different from the outflow-only velocity profile adopted by \citet{Erb_2023}.

Considering the morphology of the Voronoi-binned regions within BX710 and BX711, we construct two spatially binned \lya\ profiles from each RT model by categorizing photons into two distinct bins based on their last-scattering impact parameters: $b/R_{\rm h} \in$ (0, 0.25] and (0.25, 0.50], where $R_{\rm h}$ is the radius of the simulated halo. Here we assume that the inner halo at $b/R_{\rm h} \leq 0.5$ roughly corresponds to the high-SB Voronoi-binned regions. These spatially binned \lya\ model profiles will then be compared to pairs of observed spatially-resolved \lya\ profiles from Voronoi bins to determine the best-fit parameters\footnote{We assume that the average \lya\ profiles obtained from each Voronoi-binned region represent the \lya\ photon frequency distribution at a specific radius along a given direction. A spatially-binned model \lya\ profile is then generated for each corresponding radius and compared with the observed \lya\ profiles. Since our RT model is spherically symmetric, we fit spatially-resolved \lya\ profiles along different directions separately.}. Following the labels in Figure \ref{fig:prps}, the profile pairs\footnote{A profile pair consists of two \lya\ profiles, one originating closer to the galaxy and the other originating farther from the galaxy.} modeled are 1 $\rightarrow$ 2 and 1 $\rightarrow$ 3 for BX710, and 4 $\rightarrow$ 5, 4 $\rightarrow$ 7 and 4 $\rightarrow$ 8 for BX711\footnote{These notations indicate the relative locations of the Voronoi bins. For example, 1$\rightarrow$2 means that the inner and outer binned spectra correspond to the \lya\ profiles from Voronoi-binned regions 1 and 2, respectively.}. 

In our modeling, we have set $r_{\rm min} \rm = 1\,kpc$, $r_{\rm max} = R_{\rm h} = \rm 100\,kpc$, the clump radius $R_{\rm cl} = \rm 100\,pc$, and $\alpha = 0.15$\footnote{We note that in principle, $\alpha$ can be varied as a free parameter. However, in order to reduce the dimensionality of the parameter space, we have fixed it to 0.15 as a heuristic choice. We find that this choice yields satisfactory fits to the data and aligns with the best-fit $\alpha$ values ($\sim$ 0.2 for both BX710 and BX711) inferred from metal absorption line modeling.}. For more technical details of the \lya\ spatially-resolved modeling, we refer readers to \citet{Erb_2023}. 

\begin{figure*}[ht]
    \centering
    \includegraphics[scale = 0.85]{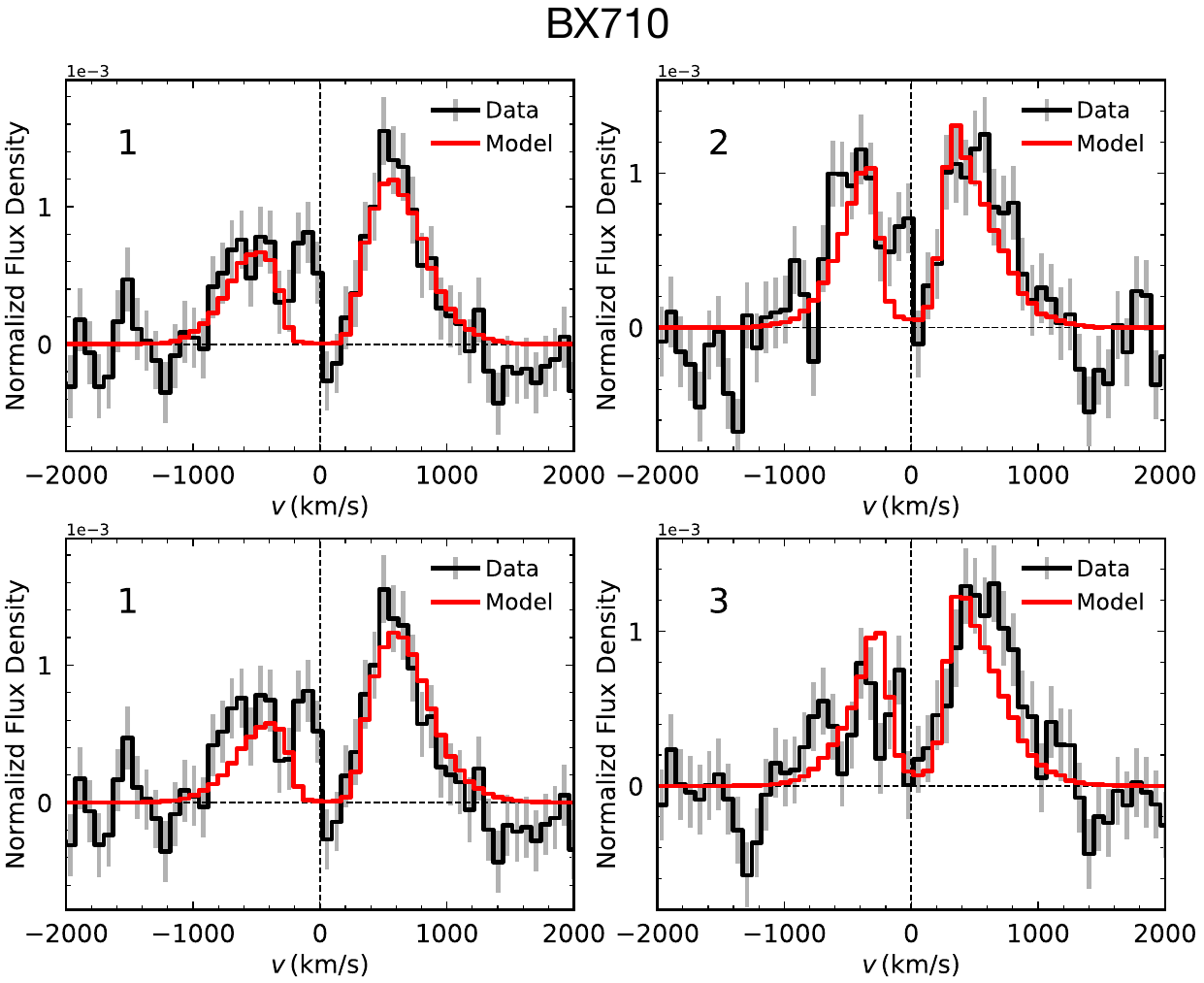}
    \caption{Spatially-resolved \lya\ RT modeling results for BX710 along two different directions (1 $\rightarrow$ 2 and 1 $\rightarrow$ 3, shown in two different rows). The observed line profiles are shown in black (with uncertainties shown in gray) and the best-fit models are shown in red. The numbers in the upper left of each panel indicate the regions defined in Figure \ref{fig:prps}. }
    \label{fig:bx710_lya}
    \end{figure*}

\begin{figure*}[ht]
    \centering
    \includegraphics[scale = 0.85]{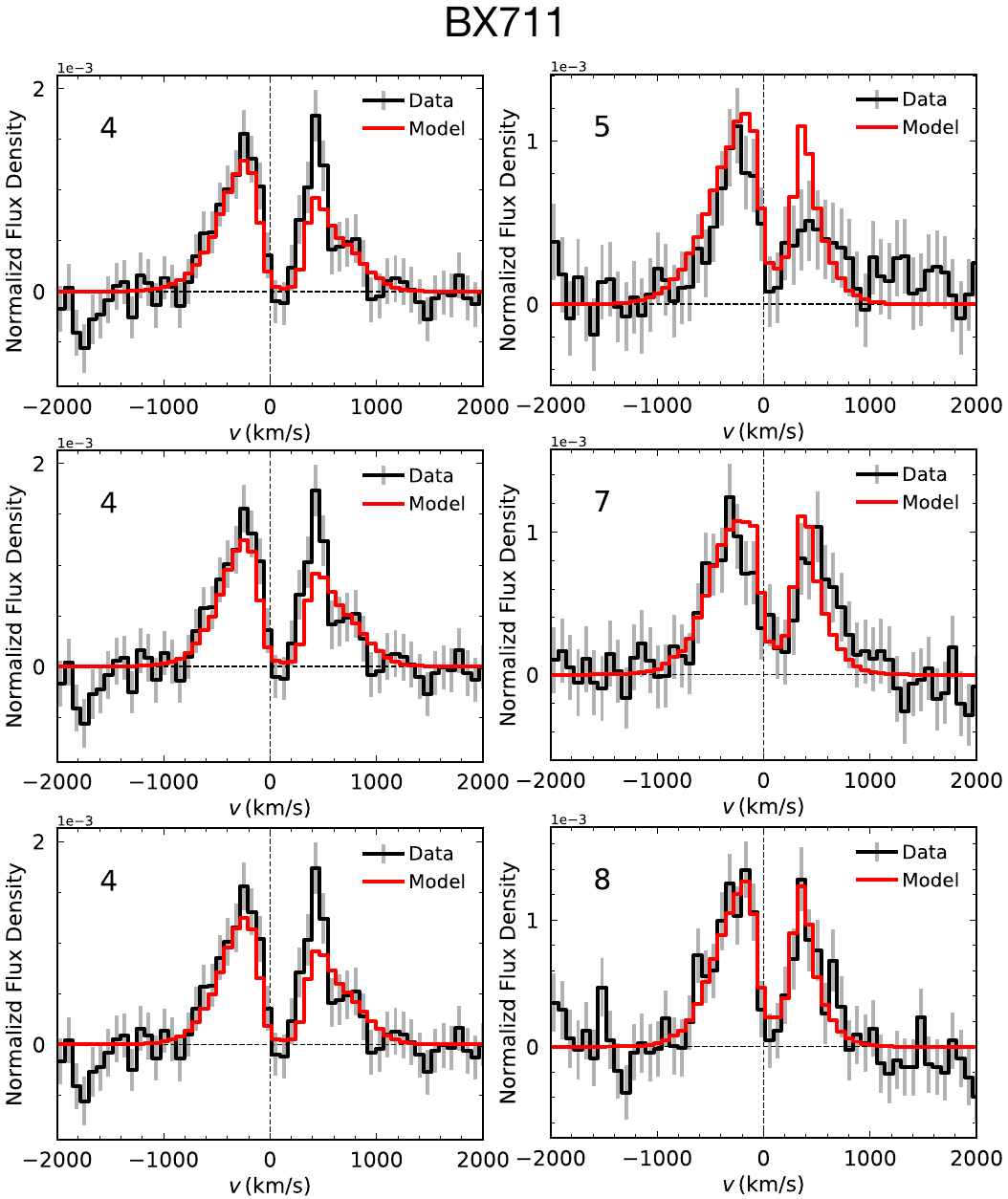}
    \caption{Spatially-resolved \lya\ RT modeling results for BX711 along three different directions (4 $\rightarrow$ 5, 4 $\rightarrow$ 7 and 4 $\rightarrow$ 8, shown in three different rows). The observed line profiles are shown in black (with uncertainties shown in gray) and the best-fit models are shown in red. The numbers in the upper left of each panel indicate the regions defined in Figure \ref{fig:prps}. }
    \label{fig:bx711_lya}
    \end{figure*}

We present the best-fits to the data in Figure \ref{fig:bx710_lya} and \ref{fig:bx711_lya}, and the best-fit parameters in Table \ref{tab:Fitting_results}.
Overall, the best-fit models are able to capture the main features of the spatially-resolved \lya\ profiles, yet there is a noticeable mismatch between the model and the data in regions 4 and 5 of BX711. Such a mismatch is primarily due to a generic property of the \lya\ profiles that emerge from a spherically-symmetric halo -- the asymmetry of the double peaks of the \lya\ profile tends to decrease with increasing impact parameter, due to a geometric effect where the component of the gas radial velocity projected in the traveling direction of the photons decreases towards the halo outskirts \citep{Erb_2023}. The discrepancy between the data and the best-fit model therefore suggests that there may be additional motion at the halo outskirts that cannot be described by Eq. (\ref{eq:radial_v}). In addition, the best-fit model parameters are not significantly direction-dependent. This is because the modeling along different directions all incorporate the average spectrum from the central region, which typically has a higher S/N ratio than the spectrum from the outer region and thus carries more weight in the fitting process.

\begin{table*}
\begin{center}
\setlength{\tabcolsep}{1pt}
\renewcommand\arraystretch{1.5}
  \scriptsize \caption{Best-fit parameters from modeling \lya\ emission and the rest-UV low ionization metal absorption lines}
  \label{tab:Fitting_results}
  \begin{tabular}{cc|ccccccc|ccccccc}
    \hline \hline 
    \multicolumn{2}{c|}{}&\multicolumn{7}{c|}{Best-fit Parameters (\lya)}&\multicolumn{7}{c}{Best-fit Parameters (LIS)}\\
    \hline 
    ID&Pair&$F_{\rm V}$&${\rm log}\,N_{\rm HI,\,{\rm cl}}$&$\sigma_{\rm cl}$&$v_{\rm 1}$&$v_{\rm 2}$&${\rm log}\,n_{\rm HI,\,{\rm ICM}}$&$\Delta v$&$A_{\rm ISM}$&${\rm log}F_{\rm V,\rm\,abs}$&$v_{\rm 1,\rm\,abs}$&$v_{\rm 2,\rm\,abs}$&$\gamma$&$\alpha$&$\sigma_{\rm rand}$\\
& &($\times 10^{-3}$) &(cm$^{-2})$&(km\,s$^{-1})$&(km\,s$^{-1})$&(km\,s$^{-1})$&(cm$^{-3})$&(km\,s$^{-1})$&&&(km\,s$^{-1})$&(km\,s$^{-1})$&&&(km\,s$^{-1})$\\
(1)&(2)&(3)&(4)&(5)&(6)&(7)&(8)&(9)&(10)&(11)&(12)&(13)&(14)&(15)&(16)\\
    \hline
Q1700{-}BX710&1 $\rightarrow$ 2 &1.4$^{+0.1}_{-0.1}$&19.5$^{+0.0}_{-0.0}$&103$^{+12}_{-9}$&173$^{+58}_{-42}$&40$^{+30}_{-24}$&-7.18$^{+0.12}_{-0.15}$&-8$^{+16}_{-15}$&0.25$^{+0.12}_{-0.10}$&-1.3$^{+0.5}_{-0.5}$&509$^{+64}_{-73}$&157$^{+43}_{-60}$&-0.4$^{+0.9}_{-0.9}$&0.2$^{+0.1}_{-0.1}$&33$^{+24}_{-23}$\\
Q1700{-}BX710&1 $\rightarrow$ 3 &1.4$^{+0.1}_{-0.1}$&19.6$^{+0.2}_{-0.1}$&87$^{+23}_{-22}$&162$^{+59}_{-39}$&26$^{+19}_{-15}$&-7.29$^{+0.17}_{-0.14}$&30$^{+35}_{-31}$\\
Q1700{-}BX711&4 $\rightarrow$ 5 &1.1$^{+0.1}_{-0.1}$&19.8$^{+0.1}_{-0.1}$&8$^{+10}_{-6}$&300$^{+24}_{-14}$&219$^{+17}_{-11}$&-7.92$^{+0.10}_{-0.06}$&120$^{+15}_{-15}$&0.12$^{+0.13}_{-0.08}$&-2.3$^{+0.2}_{-0.3}$&508$^{+62}_{-68}$&264$^{+93}_{-90}$&0.5$^{+0.4}_{-0.5}$&0.2$^{+0.1}_{-0.1}$&71$^{+36}_{-45}$\\
Q1700{-}BX711&4 $\rightarrow$ 7 &1.1$^{+0.1}_{-0.1}$&19.9$^{+0.1}_{-0.1}$&8$^{+9}_{-5}$&298$^{+22}_{-13}$&214$^{+18}_{-10}$&-7.85$^{+0.13}_{-0.10}$&127$^{+14}_{-14}$\\
Q1700{-}BX711&4 $\rightarrow$ 8 &1.1$^{+0.1}_{-0.1}$&19.8$^{+0.1}_{-0.1}$&7$^{+9}_{-5}$&299$^{+18}_{-12}$&213$^{+16}_{-8}$&-7.87$^{+0.11}_{-0.08}$&106$^{+12}_{-11}$\\
    \hline \hline
  \end{tabular}
  \end{center}
  \tablenotetext{}{\textbf{Notes.} Best-fit parameters (averages and 16\% -- 84\% quantiles, i.e., 1$\sigma$ confidence intervals) from the \lya\ and low ionization metal absorption line modeling. The columns are: (1) the object ID; (2) the pair of \lya\ spectra being modeled; (3) the clump volume filling factor; (4) the clump \HI\ column density; (5) the clump velocity dispersion; (6) the maximum clump radial outflow velocity; (7) the maximum clump radial inflow velocity; (8) the residual \HI\ number density of the ICM; (9) the velocity shift relative to the systemic redshift of the galaxy. (10) - (16) are determined from modeling the average metal absorption line profile. (10) the amplitude of absorption from the ISM; (11) the clump volume filling factor; (12) the maximum clump radial outflow velocity; (13) the maximum clump radial inflow velocity; (14) the power-law index in the clump number density profile; (15) the power-law index in the clump radial velocity profile; (16) the clump 1D velocity dispersion. }
\end{table*}

\subsection{Down-the-barrel UV absorption} 
\label{sec:abs_modeling}

We use the \texttt{ALPACA} model introduced by \citet{Li23} to model the down-the-barrel UV absorption lines of BX710 and BX711. In order to enhance the S/N ratio, we use the average absorption line profile of five different transitions: \SiII\ $\lambda$1260, 1304, 1526, \OI\ $\lambda$1302, and \CII\ $\lambda$1334. In the modeling, we adopt the physical constants and coefficients for the \SiII\ $\lambda$1260 transition.

\texttt{ALPACA} accounts for the absorption contributions from both the ISM and the CGM. The ISM absorption component is assumed to be centered around the line center, characterized by an amplitude of $A_{\rm ISM}$, and a standard deviation of $\sigma_{\rm ISM}$. In the CGM, the absorbing gas clumps are described by their total volume filling factor $F_{\rm V}$, radial velocity profile, and number density profile. For the sake of simplicity, we assume that all the clumps have a constant radius of $R_{\rm cl} = 100\,{\rm pc}$ and an ion column density of $N_{\rm ion,\,cl} = 10^{15}\,{\rm cm^{-2}}$.

The radial motion of the clumps in the CGM has two components: random motion characterized by the velocity dispersion $\sigma_{\rm rand}$, and a radial outflow / inflow. To ensure consistency, we adopt the radial velocity profile given by Equation \ref{eq:radial_v}, while treating $v_1, v_2\ {\rm and}\ \alpha$ all as free parameters. The clump number density, which is proportional to the gas geometric covering fraction, is assumed to vary radially as a power-law with a power-law index of $-\gamma$:
\begin{equation}
n_{\rm cl}(r) = n_{\rm cl,\,0} \Big(\frac{r}{r_{\rm min}}\Big)^{-\gamma}
\label{eq:n_cl}
\end{equation}
where $n_{\rm cl,\,0}$ is the clump number density at $r_{\rm min}$. 

\begin{figure*}[ht]
    \centering
    \includegraphics[scale = 0.5]{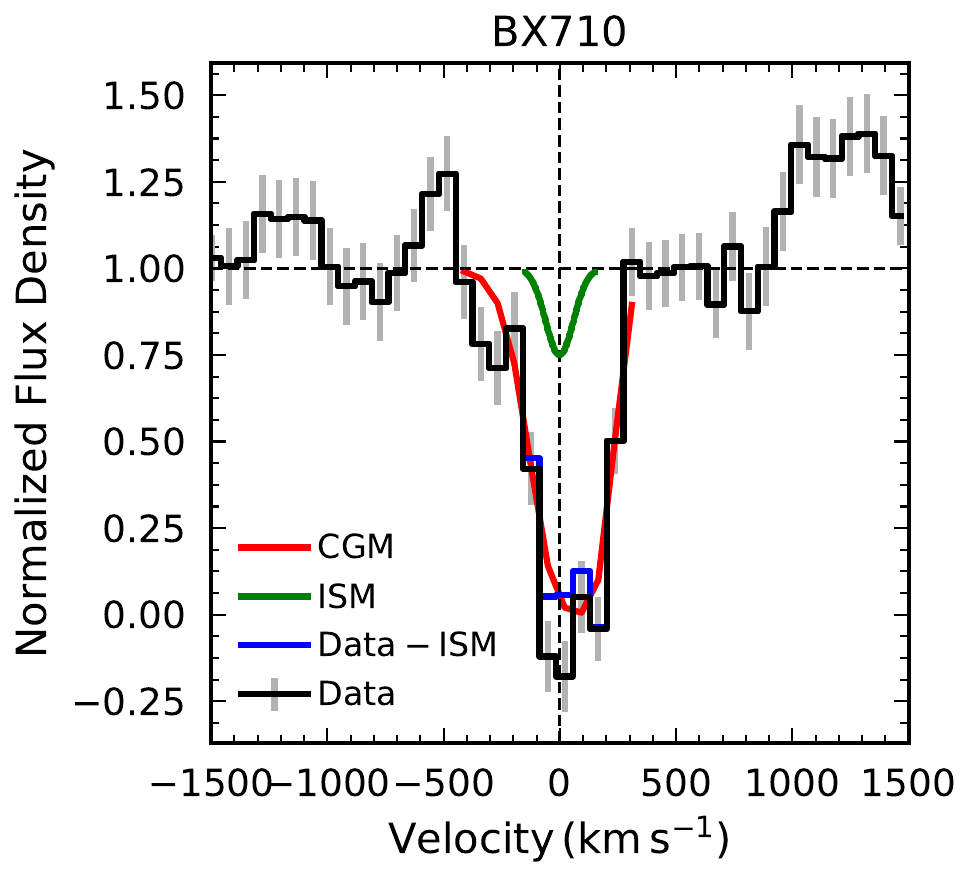}
    \includegraphics[scale = 0.5]{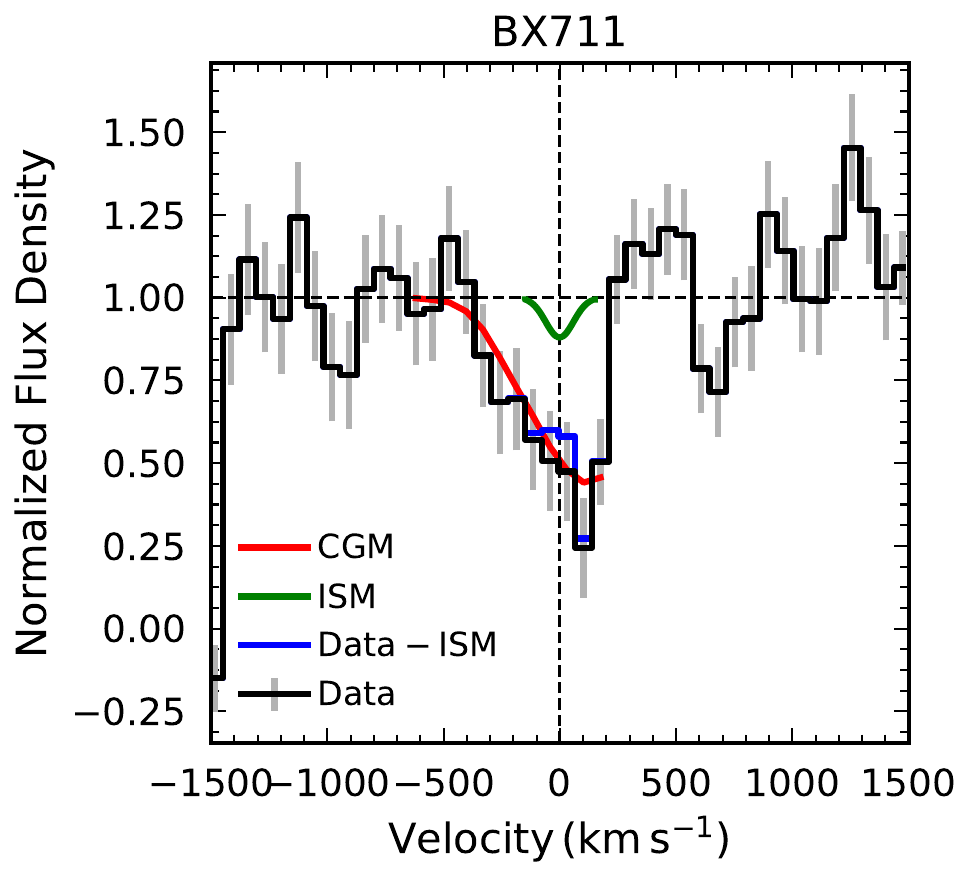}
    \caption{Results of modeling the average metal absorption line profile for BX710 and BX711. The data are shown in black and the best-fit models are shown in green (the ISM component) and red (the CGM component), respectively. The ``data - ISM'' profile is also shown for comparison to the best-fit CGM model.}
    \label{fig:abs_best_fits}
    \end{figure*}

We present the best-fits to the data in Figure \ref{fig:abs_best_fits}, and the best-fit parameters in Table \ref{tab:Fitting_results}.

\section{Discussion} \label{sec:dis}

\textbf{In} this section, we compare BX710 and BX711's UV spectral features to those of UV-continuum selected galaxies and LAEs at $z \sim 1-3$ as well as other epochs. Additionally, we explore the implications of our best-fit \lya\ RT models and radially varying velocity profiles for BX710 and BX711's spectra in the context of their environments and explore potential scenarios that could be responsible for the unusual features we observe in their spectra.

\subsection{\lya\ Profiles}

BX710's \lya\ profiles exhibit blue-to-red peak ratios ranging from $\sim0.5-0.8$, which are consistent with most UV continuum-selected galaxies and LAEs whose spectra feature double-peaked \lya\ emission at $z\sim2-3$ \citep{Trainor_2015, Erb_2018, Matthee_2021, Erb_2023}. Green pea galaxies at $z\sim0.1-0.3$ \citep[e.g.,][]{Henry_2015, Yang_2016, Orlitova_2018}, LAEs at $z\sim3-6$ \citep{Kerutt_2022}, and an LAE at $z=6.6$ \citep{Matthee_2018} which have double-peaked \lya\ emission also demonstrate comparable \lya\ peak ratios to BX710. BX711's \lya\ profiles, conversely, possess blue to red peak ratios of $\sim 0.7-1.6$. Blue to red \lya\ peak ratios greater than unity have seldom been reported for any object, which is discussed further in Section \ref{sec:intro}. 

The peak separations of the \lya\ profiles in BX710's \lya\ halo are unusually wide, spanning a range of $\sim$ 900-1000 km s$^{-1}$, in comparison to UV continuum-selected galaxies and LAEs with double-peaked \lya\ emission at z$\sim$2-3, which exhibit peak separations of $\sim$200-900 km s$^{-1}$ \citep{Kulas_2012, Trainor_2015, Berg_2018, Erb_2018, Naidu_2021, Erb_2023, Mukherjee_2023}. These high peak separations likely result from a relatively high column density \citep{Dijkstra_2014, Verhamme_2015, Erb_2023} of the HI clumps in the CGM. The blue \lya\ peaks in BX710's \lya\ halo are also exceptionally broad and exhibit more than one peak in all three binned regions, which implies that these blue peaks may have multiple components. As a result, our calculations of BX710's \lya\ peak separations may be oversimplified. Meanwhile, the \lya\ profiles in BX711's \lya\ halo exhibit peak separations that are wide at $\sim$ 700-900 km s$^{-1}$ but still consistent with the aforementioned peak separations of LAEs and UV-continuum selected galaxies at this epoch \citep{Kulas_2012, Trainor_2015, Erb_2023, Mukherjee_2023}.

\subsection{Radiative Transfer Models}
\label{subsec:rt_models}

As shown in Figures \ref{fig:bx710_lya}, \ref{fig:bx711_lya} and \ref{fig:abs_best_fits}, our models have reproduced the \lya\ and LIS line profiles reasonably well. Both \lya\ and LIS modeling suggest that a transition in the clump radial velocity profile from outflow to inflow is capable of explaining the observational data. However, there is a noticeable discrepancy in the best-fit values of clump radial velocity. Specifically, the maximum outflow and inflow velocities inferred from LIS line modeling are both larger (approximately $\sim$ 500\,km s$^{-1}$ and --200\,km s$^{-1}$, respectively) than those inferred from \lya\ modeling for BX710 (and to a lesser extent for BX711).

In general, the gas radial velocities play different roles in determining the characteristics of the observed line profiles for \lya\ emission and LIS metal absorption lines. Specifically, in the spatially-resolved \lya\ modeling, the clump radial outflow or inflow velocity is mainly constrained by the flux ratio of the double peaks in the \lya\ profiles, whereas in LIS line modeling, the clump radial outflow or inflow velocity essentially corresponds to the range of velocities where significant absorption is observed. The disparities in the inferred gas radial velocities may therefore arise from additional gas motion (either radial or non-radial) not fully captured by our kinematic model, which could affect the line shape of \lya\ emission more significantly than the LIS metal absorption. Another possible reason is the intrinsic difference between the LIS line profiles and \lya\ profiles---the LIS line profiles are down-the-barrel observations that trace the underlying properties of the gas along our line-of-sight, whereas the spatially-resolved \lya\ profiles are extracted from regions at different impact parameters, encompassing a much broader range of physical space especially considering the resonant scattering of \lya\ photons. 

In addition to the clump radial velocity, several other clump parameters derived from \lya\ and LIS line modeling are not fully consistent either, including the clump volume filling factor ($F_{\rm V}$) and the clump radial distribution (characterized by the power-law index $\gamma$, which is $\sim$ 2 in the \lya\ modeling\footnote{We assume a power-law distribution for the clump number density $n_{\rm cl}(r) \propto r^{-2}$ in the RT model, as the power-law index cannot be varied continuously as in \texttt{ALPACA}.}). These discrepancies may also stem from the fact that LIS line profiles and \lya\ emission line profiles probe different regions within the halo. The apparent asymmetry in the radial variation of the emission may also contribute to these discrepancies.

\subsection{Accretion Scenarios}

Below we discuss multiple accretion scenarios which may explain the gas flowing into BX710 and BX711. We note that we cannot rule out any of the following scenarios without further observations, which we discuss in Section \ref{subsec:future}. Additionally, any combination of these scenarios may be occurring in tandem.

\subsubsection{Cosmic Web}

As we described in Section \ref{ssec:fil}, BX710 and BX711 are located in the complex environment of the HS1700+64 protocluster, which is composed of numerous \lya\ blobs, \lya\ emitters, and UV-continuum selected galaxies. Within this protocluster, we delineated two nearly linear, intersecting filaments containing 34 total UV-continuum selected galaxies, one of which contains BX710 and BX711. 

As observed in \citet{Umehata2019}, there may be cool HI gas flowing along these large-scale filaments which subsequently accretes onto BX710 and BX711, especially since BX710 and BX711's \lya\ surface brightness contours are roughly aligned with one of these filaments. As a result, the stronger blue peaks which we observe in BX711 and the inflows which our models determined for BX710 and BX711 may very well be due to cool HI gas from the cosmic web accreting onto them. 

We caution, however, that this scenario could only occur if the the IGM gas is already metal-enriched or if cool HI gas from the cosmic web sweeps metal-enriched CGM gas along with it as it accretes, since BX710 and BX711's redshifted UV absorption features and our best-fit velocity models to them demonstrate that metal-enriched gas is accreting onto both galaxies. Additionally, accretion of metal-poor gas from the cosmic web could be less likely to stimulate growth in BX710 in particular because cosmological simulations suggest that galaxies with log(M$_{*}$) $\gtrsim 10.5$ at cosmic noon tend to shock heat cool \HI\ gas as it accretes, which prevents this gas from forming new stars until it cools \citep{Birnboim_Dekel_2003, Dekel_2006}.

\subsubsection{Galaxy Interactions}

Several factors suggest that BX710 and BX711 may very well be interacting, including this galaxy pair's small projected separation of 48 pkpc and $\Delta v \approx 30$ km s$^{-1}$ \citep{Ventou_2019}; the tidal \Ha\ tail observed by \citet{Law_2009} stretching northwest from BX710; the offset between BX710's strongest \lya\ emission and UV continuum emission in the opposite direction from BX711; and the \lya\ bridge connecting BX710 and BX711's \lya\ halos. \citet{Solimano_2022} observed a similar \lya\ bridge between the \lya\ halos of a lensed, potentially merging galaxy pair at z = 2.9, which are separated by a distance of only $d_{UV} \sim$ 10 pkpc in their source plane. However, the \lya\ bridge in \citet{Solimano_2022} was far more extensive, causing the individual \lya\ halos for each galaxy to form a single \lya\ halo surrounding the entire galaxy pair, much like a \lya\ blob.

To make matters more complicated, only 10 pkpc east of BX711 there is another UV-continuum selected galaxy, CS7, which is labeled and denoted by a red circle in the \textit{HST} image displayed in the bottom panel of Figure \ref{fig:im_lya}. As mentioned in Section \ref{subsec:targetsandfriends}, the redshift of this object is unfortunately not well constrained, thereby producing a tentative separation from BX711 of $\Delta v \approx 600$ km s$^{-1}$. If CS7 is indeed located in such close proximity to BX711, the two galaxies could be interacting, especially considering CS7's stellar mass is $\sim 30$ times greater than that of BX711, making it very similar to BX710's stellar mass. Moreover, even though CS7's spectrum exhibits little \lya\ emission of its own at its estimated redshift, CS7 happens to coincide with Voronoi bin 7 in BX711's \lya\ halo, which has strong \lya\ emission and the highest blue-to-red \lya\ peak ratio. Contributions of \lya\ emission from CS7 may partially explain the high blue-to-red peak flux ratio in this bin. 

Additionally, we note that BX710 and BX711's CGMs likely overlap due to the proximity of these galaxies. As a result, the gas in their CGMs is likely interacting and disrupting each CGM's kinematics as a result. The interference between their CGMs may explain the difficulty that the radiative transfer models experienced with reproducing the \lya\ profiles from the Voronoi-binned regions, which are heavily dependent on the gas kinematics in these galaxies' CGMs, and could be a cause of the non-radial motions that the models do not appear to capture, as is discussed in Section \ref{subsec:rt_models}\textbf{.}

\subsubsection{Recycled Accretion}

Aside from BX710 and BX711, redshifted metal absorption lines have 
previously been reported several times in the spectra of lower redshift ($z \lesssim 2$) and higher redshift ($z\gtrsim 2$) galaxies \citep[e.g.][]{Giavalisco_2011, 2012ApJ...760..127M, Rubin_2012, Calabro_2022, Weldon_2023}. In the KBSS sample, for instance, 86 out of 907 (or nearly 10\% of) sampled galaxies with redshifts from both nebular emission lines ($z_{neb}$) and interstellar absorption lines ($z_{IS}$) exhibit $z_{IS} > z_{neb}$, or net redshifted interstellar absorption features, although the culpable accretion mechanisms have not been constrained. In other studies, recycled accretion, during which metal-enriched gas flowing out from a galaxy reaccretes onto it after failing to escape its CGM, has often been considered a prime suspect for these redshifted absorption lines.

\citet{Rubin_2012}, for example, attributed the redshifted Mg II and Fe II absorption lines which they detected in $z \lesssim 1$ galaxies to either accretion of recycled gas from the CGM or metal-enriched gas from interactions with satellite galaxies. \citet{2012ApJ...760..127M} also reported redshifted UV absorption lines in galaxies at $z \sim 1$, and noted that these lines may have arisen from recycled accretion alone or recycled accretion driven by IGM accretion. Similarly, at $z \sim 2.30$, \citet{Weldon_2023} found redshifted low-ionization UV features in the spectra of three SFGs and concluded that these spectral features likely resulted from recycled accretion as opposed to accretion from the cosmic web.

As these studies of galaxies with redshifted absorption lines have shown, it can be difficult to distinguish whether these features are formed by recycled accretion, other metal-enriched accretion scenarios such as galaxy interactions, or even a combination of the two. In fact, hydrodynamical simulations of merging galaxies predict that a sizeable fraction of the gas galaxies accrete during a merger originates from the CGM of these galaxies \citep{Sparre_2021}. Even in the presence of clear evidence of a merger, it may be challenging to distinguish this accreting CGM gas from recycling CGM gas. Bearing these considerations in mind, we infer that BX710 and BX711's redshifted low-ionization absorption features signify that the gas flowing into them is at least slightly metal-enriched and may stem from recycled accretion and/or any of the other aforementioned accretion scenarios.

\subsection{Future Prospects}
\label{subsec:future}

Prospectively, high resolution imaging and further spectroscopic observations will be necessary to help resolve remaining questions about the kinematics and sources of the gas flowing into BX710 and BX711.
Specifically, spatially resolved \Ha\ observations of BX711, CS7, and CS10 would
help constrain the likelihood of galaxy interactions among the galaxy pair and nearby galaxies; such observations have already been undertaken for BX710 \citep{Law_2009}. Likewise, higher S/N spectroscopic observations of the nearby objects CS7 and CS10 will be necessary to confirm their redshifts, and consequently, their proximity to BX710 and BX711. 

Additionally, the discrepancies between the best-fit parameters of the \lya\ and UV absorption line models, which is discussed above in Section \ref{subsec:rt_models}, and the inability of the models to match the peak ratio trends in BX711's \lya\ halo suggest that comparisons with more complex models which do not assume spherical symmetry, such as hydrodynamical simulations with \lya\ radiative transfer, may elucidate the geometry and kinematics of this galaxy pair. Further, higher S/N observations of BX710, BX711, and the other targets in the field would better inform these simulations of the galaxies' physical conditions.  

\section{Summary}\label{sec:summary}
Here we have investigated KCWI observations of a UV-continuum selected galaxy pair, Q1700-BX710 and Q1700-BX711, in the HS1700+64 protocluster at $z = 2.3$ to ascertain whether gas is falling into these galaxies from a galaxy merger or another environmental source, as signaled by the blue-dominated \lya\ emission in BX711's spectrum. Our findings are summarized as follows:

\begin{enumerate}
    \item From least-squares elliptical isophote fits to BX710 and BX711's \lya\ halos, we find that BX710 and BX711's \lya\ halos are elongated and oriented at $22.9 \pm 2.3 \degree$ east of north and $29.0 \pm 2.4 \degree$ east of north, respectively. Through least squares linear regression, we determine that a group of thirteen UV continuum-selected galaxies in the HS1700+64 protocluster, including BX710 and BX711, form an approximately linear filament oriented at $31.2 \pm 3.6 \degree$ east of north and around 1.48 comoving Mpc wide. We therefore conclude that BX710 and BX711's \lya\ halos are aligned with the galaxy filament within a 2$\sigma$ and 1$\sigma$ uncertainty, respectively.
    
    \item The KCWI spectra we extracted from Voronoi-binned regions in BX711's \lya\ halo all, except in the case of one outer bin, feature double-peaked \lya\ emission with stronger blueshifted peaks and large ($\gtrsim 700$ km s$^{-1}$) \lya\ peak separations that span a range of $\sim$150 kms$^{-1}$, suggesting the presence of inflowing gas throughout BX711's CGM. Meanwhile, the majority of the binned regions in BX710's \lya\ halo have double-peaked \lya\ profiles with stronger redshifted \lya\ peaks and notably broad \lya\ peak separations of up to $\sim1000$ km s$^{-1}$ with a range of $\sim$150 kms$^{-1}$. 
    
    \item The low-ionization metal UV absorption features in BX710 and BX711's 1D optimally extracted spectra are offset from the systemic velocity on average by $111 \pm 12$ km s$^{-1}$ and $31 \pm 25$ km s$^{-1}$, respectively. As opposed to the redshifted absorption features we observe in the galaxy pair's spectra (albeit only slightly redshifted in the case of BX711), which can be associated with gas inflows, blueshifted absorption lines caused by outflowing gas are more commonly observed among $z\sim2-3$ galaxies \citep{2003ApJ...588...65S, Jones_2012, Steidel_2010}.
    
    \item We applied radiative transfer models to the \lya\ profiles from the  Voronoi-binned regions which assume that the CGM contains HI clumps that gradually switch from outflowing to inflowing with increasing galactocentric radius. Additionally, we fit the average low-ionization UV absorption line profile from BX710 and BX711's spectra with the same radially varying velocity profile. The best-fit models to BX710 and BX711's \lya\ and UV profiles confirm that gas is flowing into BX710 and BX711's \lya\ halos.

    \item We explore various accretion scenarios that may account for the gas flowing into BX710 and BX711. Given this galaxy pair's redshifted, low-ionization metal UV absorption features, we find that any metal-enriched accretion scenario, such as recycled accretion or metal-enriched gas accreting from the cosmic web, could be at play.

\end{enumerate}

    Investigations of galaxies with blue-dominated \lya\ emission and redshifted absorption features remain rare in the literature, likely due to the rarity with which they have been detected. In the KBSS sample alone, only three galaxies exhibit both redshifted interstellar absorption features and blue-dominated \lya\ emission out of 907 total sample galaxies with redshifts determined from interstellar absorption features and nebular emision lines. To the best of our knowledge, there are currently no IFU studies of galaxies with blue-dominated \lya\ emission and redshifted absorption lines; however, one such galaxy, the SFG J1316+2614 reported by \citet{Marques_Chaves_2022}, has been studied spectroscopically. Although detections of UV-continuum selected galaxies and LAEs with just blue-dominated \lya\ emission are substantially more prevalent \citep{Kulas_2012, Trainor_2015, Berg_2018, Marques_Chaves_2022, Furtak_2022, Guo_2023, Mukherjee_2023}, very few of these galaxies have been examined by IFU observations either \citep{Furtak_2022, Mukherjee_2023}, and the environments of those outside the KBSS sample remain largely unconstrained.
    
    Despite the rarity of galaxies with blue-dominated \lya\ emission and redshifted absorption features, these galaxies are especially important to study, as they display multiple signatures of gas accretion which can help constrain the kinematics of the accreting gas. We encourage the undertaking of extensive integral field observations of galaxies with blue-dominated \lya\ profiles and redshifted UV absorption lines, if available, as well as careful studies of their environments in order to clarify the relative contributions of environmental sources to galaxy growth at cosmic noon and earlier times.

\begin{acknowledgements}
We thank the anonymous referee for their insightful comments, which have enhanced the clarity of this work. C.B was supported by the University of Wisconsin-Milwaukee's Office of Undergraduate Research through the Support for Undergraduate Research Fellows (SURF) Award and Senior Excellence in Research Award (SERA). C.C.S. and Z.L. were supported in part by U.S. National Science Foundation grant AST-2009278 and by the JPL/Caltech President's and Director's Fund. D.K.E. was supported by the US National Science Foundation (NSF) through Astronomy \& Astrophysics grant AST-1909198. The authors wish to recognize and acknowledge the very significant cultural role and reverence that the summit of Maunakea has always had within the indigenous Hawaiian community. We are deeply grateful for the opportunity to conduct observations from this mountain. Authors C.B. and D.K.E. acknowledge in Milwaukee that we are on traditional Potawatomi, Ho-Chunk and Menominee homeland along the southwest shores of Michigami, North America's largest system of freshwater lakes, where the Milwaukee, Menominee and Kinnickinnic rivers meet and the people of Wisconsin's sovereign Anishinaabe, Ho-Chunk, Menominee, Oneida and Mohican nations remain present. We are very grateful for the opportunity to do research in the homelands of these sovereign nations. 
\end{acknowledgements}

\bibliography{Q1700_BX710_BX711}
\bibliographystyle{aasjournal}

\end{CJK*}
\end{document}